\documentclass[twocolumn]{revtex4}
\usepackage{helvet,times}
\usepackage{bm,textcomp}
\usepackage{bbm}
\usepackage{amsmath}
 \usepackage{graphicx}

\usepackage[sort&compress]{natbib}

\def\dd{\mbox{d}}

\begin{document}

\setcounter{page}{1} %first page number

\title{The Effects of Statistical Multiplicity of Infection on Virus
  Quantification and Infectivity Assays}

%\author{Bhaven Mistry, Maria-Rita D'Orsogna, Tom Chou}
\author{Bhaven Mistry$^{1}$, Maria R. D'Orsogna$^{1,2}$, and Tom Chou$^{1,3}$ \\
$^{1}$Department of Biomathematics, University of California, Los Angeles, CA 90095-1766 \\
$^{2}$Department of Mathematics, California
State University, Northridge, CA 91330 \\
$^{3}$Department of
Mathematics, University of California, Los Angeles, CA 90095-1555}

% generate the title page from the info in the headers above
%\pagebreak[4]
%\maketitle

%Abstract environment needs 3 arguments. They are
%1. The abstract
%2. Received date
%3. Address, email

\begin{abstract}%
Many biological assays are employed in virology to quantify parameters
of interest. Two such classes of assays, virus quantification assays
(VQA) and infectivity assays (IA), aim to estimate the number of
viruses present in a solution, and the ability of a viral strain to
successfully infect a host cell, respectively. VQAs operate at
extremely dilute concentrations and results can be subject to
stochastic variability in virus-cell interactions. At the other
extreme, high viral particle concentrations are used in IAs, resulting
in large numbers of viruses infecting each cell, enough for measurable
change in total transcription activity. Furthermore, host cells can be
infected at any concentration regime by multiple particles, resulting
in a statistical multiplicity of infection (SMOI) and yielding
potentially significant variability in the assay signal and parameter
estimates. We develop probabilistic models for SMOI at low and high
viral particle concentration limits and apply them to the plaque
(VQA), endpoint dilution (VQA), and luciferase reporter (IA) assays. A
web-based tool implementing our models and analysis is also developed
and presented.  We test our proposed new methods for inferring
experimental parameters from data using numerical simulations and show
improvement on existing procedures in all limits.
%
% {}Insert Received for publication Date and in final form Date.  
% {}Insert Corresponding address and emails
\end{abstract}

\maketitle %%The above information typeset through this command

\section{Introduction}
Understanding viral dynamics is an important task in medicine,
epidemiology, public health, and, in particular, for the development
of antiviral therapies and vaccines.  Drugs that hinder viral
infection include blockers of viral entry into the host cell
\cite{Pegu2014, Osbourn1998,Su_Qiu_2012,Platt1,Platt2005,Fatkenheuer}
and inhibitors of genetic activity and protein assembly inside the
cytoplasm and nucleus
\cite{Jonckheere2000,thierry2015integrase,Paterson2000}. Mechanistic
models of drug action have recently emerged as useful tools in helping
design ad-hoc experiments to study drug efficacy and in interpreting
results \cite{Wilen2012,Qian2009,Boulant,Chou2}. Mathematical models
typically assume prior knowledge of given physical quantities
pertaining to the virus, host cell, or the biological assay being
studied. Once these parameters are assigned, viral and cell population
dynamics and their statistical properties can be predicted.  Among the
different experimental assays, one often seeks to evaluate the number
of virus particles in a stock solution or the number of viruses that
have successfully infected host cells
\cite{Chikere201381,Johnston,Webb2016, Killian2008, Mascola2002,
  Brown2014,Fatkenheuer}.

In the case of virus quantification assays (VQA), performing repeated
controlled experiments on viral dynamics or comparing results across
multiple studies requires knowing how many viruses are present in the
initial stock solution of each experiment \cite{Platt1,Platt2005}.
Furthermore, antigens that induce immune responses against viral
infections may be engineered from viral components such as
capsid proteins, viral enzymes, and genetic vectors \cite{Sette2003},
and may be used in the development of vaccines. Being able to
determine the exact number of virus-derived antigens helps control the
efficacy of vaccines and optimize yield
\cite{Gerdil2003,Tree2001,Neumann2005}.

Given the central role of VQAs, several assays have been designed to
estimate viral particle counts. These include plaque
\cite{Kropinski2009} and endpoint dilution
\cite{Johnson1990,Neumann2005} assays, which will be discussed in more
detail in the remainder of this work. For now, we note that these
assays involve repeatedly diluting an initial solution of virus
particles in the presence of a layer of plated cells, until viral
concentrations are low enough that the dynamics of an individual virus
can be extrapolated.  At these low particle counts, however, the
discrete nature of the infection process cannot be neglected and can
cause substantial discrepancies when replicating experiments. Average
quantities are not necessarily representative, and a more in-depth
approach in quantifying virus-cell interactions is necessary.
 
Infectivity assays (IA), on the other hand, aim to quantify the number
of viruses that have successfully infected host cells under varying
antiviral drug environments \cite{Chikere201381,Webb2016,Johnston}.
IAs may measure the total transcription activity across all cells,
such as the luciferase reporter assay \cite{Johnston, Agrawal2009}, or
may count the number of host cells that were successfully infected,
such as the enzyme-linked immunosorbent assay (ELISA) and the
immunofluorence assay with fluorescence activated cell sorting (FACS)
\cite{Johnston,Mascola,Chikere201381, Platt1}.  These assays are
performed using undiluted solutions with large numbers of viral
particles, reducing stochastic variability. The average number of
viruses that infect a cell is estimated as the ratio of the number of
viruses in solution to the number of plated cells, a quantity known as
the multiplicity of infection (MOI) \cite{Brown2014}. However, each
cell may be infected by different numbers of viruses distributed around
the average given by the MOI. In these cases, one may be interested in
the complete probability distribution for the number of virus
infections in each plated cell and in the related statistical
variance.

In this paper, we derive a probability model for the distribution of
viral infections per host cell, which we call the statistical
multiplicity of infection (SMOI).  The SMOI can be used as a starting
point to help estimate the number of viral particles in solution in
VQAs, and to determine a viral strain's ability to successfully infect
host cells in IAs.  In Section~\ref{SEC_PROBABILITY}, we present the
mathematical foundations for the SMOI in the two experimentally
relevant parameter regimes of small and large viral particle counts
and derive a probability model for the total number of infected cells
under any dilution level. In Section \ref{SEC_PLAQUE}, we apply our
models to the plaque assay and formulate a new method of analyzing
plaque count data. In Section~\ref{SEC_ENDPOINT} we employ a special
case of the derived probability distribution to the endpoint dilution
assays and compare our results to those arising from traditional
titration techniques such as the Reed and Muench \cite{Reed1938} and
Spearman-Karber methods \cite{Hamilton1977}. In Section
\ref{SEC_LUCIFERASE}, we use the large particle limit of our model to
describe the luciferase reporter assay. Lastly, a discussion of our
results, a side-by-side comparison with existing methods, and a link
to web-based data analysis tools are provided in Section
\ref{SEC_DISC}. Mathematical appendices and further discussion of
experimental attributes such as cell size variability, coinfection,
viral interference, and optimal experimental design using parameter
sensitivity analysis are presented in the Supplemental Information
(SI).

%%%%%%%%%%%%%%%%%%%%%%%%%%%%%%%%%%%%%%%%%%%%%%%%%%%%%%%
\section{Methods} 

\begin{figure}[t!]
\includegraphics[width=1.7in]{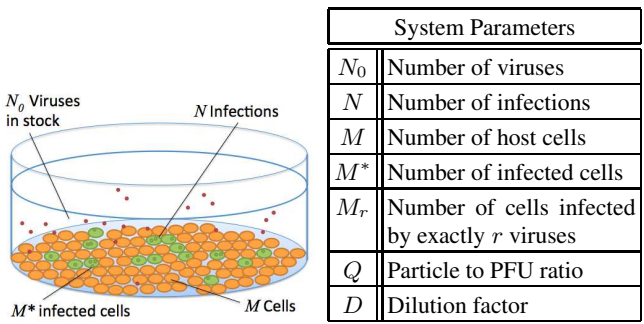}
%\:\hspace{3in}
\begin{tabular}[b]{ |c||p{3.3cm}| }
 \hline
 \multicolumn{2}{|c|}{System Parameters} \\
 \hline
 \hline
 $N_0$   & Number of viruses\\
 \hline
 $N$   & Number of infections\\
 \hline
 $M$   & Number of host cells\\
 \hline
 $M^*$ & Number of infected cells\\
 \hline
 $M_r$ & Number of cells infected by exactly $r$ viruses \\
 \hline
 $Q$ & Particle to PFU ratio \\
 \hline
 $D$ & Dilution factor \\
% \hline
% $T$ & Number of assay trials \\
% \hline
% $P_{d,t}$ & Plaque assay data \\
% \hline
% $E_d$ & Endpoint dilution data \\
% \hline
% $L_{t}^{\mathrm{data}}$ & Luciferase reporter assay data \\
 \hline
\end{tabular}
\caption{A typical assay includes a plate of $M$
  host cells inoculated with a solution of $N_0$ viruses. Each viral
  particle has some probability of infection and the total number $N$
  of infections are distributed to the $M^*$ infected cells. The
  probability of infection is roughly estimated with the reciprocal of
  the \textit{a priori} measured particle to PFU ratio $Q$.\label{FIG_PLATEDCELLS}}
\end{figure}

%------------------------------------------------------------------------------------------------------------------------------------------------
\subsection{Probabilistic Models of Statistical Multiplicity of Infection (SMOI)} 
\label{SEC_PROBABILITY}

A typical viral assay is initiated by laying a monolayer of $M$ cells
on the bottom of a microtiter well, as illustrated in
Fig.~\ref{FIG_PLATEDCELLS}
\cite{Johnson1990,Kropinski2009,Killian2008}. Although variability
exists among experiments, $M$ is often set within the range of
$10^4$--$10^5$ \cite{Agrawal2009,Chikere201381} and is assumed to be a
known experimental parameter. A supernatant containing $N_0$ virus
particles in the range of $10^5$--$10^7$
\cite{Agrawal2009,Kropinski2009,Johnson1990}, is then added to the
microtiter well. While, theoretically, all $N_0$ particles are capable
of infection, not all will successfully infect a cell. Since infection
of a host cell requires a complex sequence of biochemical processes
that may include receptor binding, membrane fusion, reverse
transcription, nuclear pore transport, and DNA integration
\cite{Wilen2012,Brown2014}, virus particles that fail at one or
several of these sequential steps lead to abortive infections. To
differentiate, the particles that do succeed are called infectious
units (IU) or plaque forming units (PFU). We will denote the number of
IUs as $N\leq N_0$. Depending on the strain of virus, the particular
experimental protocol used, and specific conditions of the assay, the
random quantity $N$ is distributed according to $N_0$ and the overall
effective probability that an arbitrary viral particle successfully
infects a host cell. A proxy that is typically used in place of this
effective probability is the ``particle to PFU ratio'' $Q$, an
experimentally determined parameter that quantifies, on average, the
minimum number of particles required to ensure at least one infected
cell \cite{SCHWERDT1957,Klasse2015}. $Q$ is often treated as an
\textit{a priori} measured quantity, primarily associated with the
particular strain of virus being studied. Low values of $Q$, such as
with poliovirus ($Q=30$) \cite{Klasse2015}, have a high likelihood of
successful infection compared to viruses with large $Q$, such as HIV-1
($Q=10^7$) \cite{Layne1992}. Thus, the reciprocal $Q^{-1}$ can be
interpreted as the probability for a single virus to infect a host
cell.  Assuming an initial stock of $N_0$ particles, the discrete
probability density function of $N$ is
\begin{equation}\label{EQ_N_BINOM_PROB}
\mathrm{Pr}\left(N = n\vert N_0, Q\right) = \binom{N_0}{n} 
\left(Q^{-1}\right)^n \left(1 - Q^{-1}\right)^{N_0 - n},
\end{equation}
which defines a binomial distribution with parameters $N_0$ and
$Q^{-1}$. Although we assume $Q$ to be \textit{a priori} known, in
actuality, the probability of a virus successfully infecting a host is
highly dependent on the methods used to harvest the virus stock, the
experimental parameters of the assay, the host receptor concentrations
and binding rates, and the dynamics of the physiological processes
leading to infection \cite{SCHWERDT1957,Turner2004}. A thorough
investigation into these processes would be necessary to
mechanistically model $Q$ and is outside of the scope of this
paper. However, we will discuss in Section~\ref{SEC_DISC} how, with
direct measurements of certain other parameters, especially $N_{0}$,
our derived methods may also be used to infer $Q$.

\begin{figure}[t!]
\begin{center}
\includegraphics[width=3.5in]{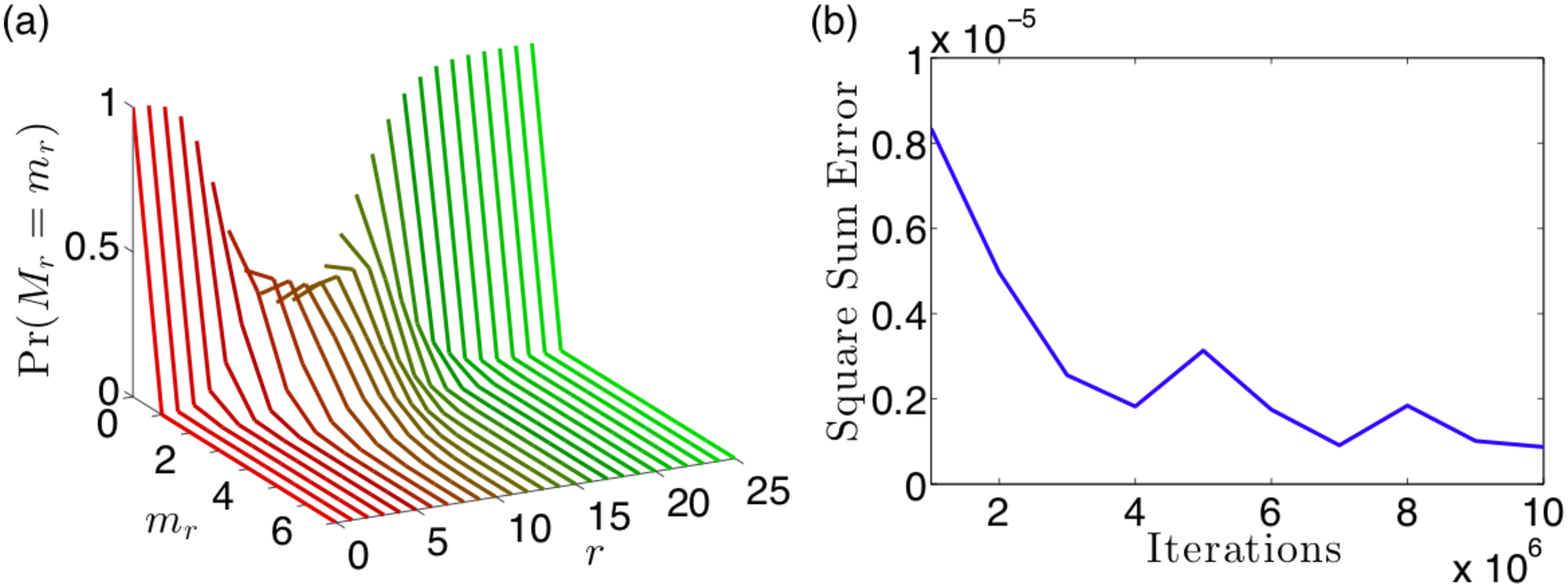}
\caption{(a) A collection of curves of the
  probability of finding $m_r$ cells that have been infected by
  exactly $r$ IUs given a total number of IUs $N=100$ and a total
  number of cells $M=10$ using Eq.~\ref{EQ_PROB}. With $N/M=10$, we
  expect very few cells to be uninfected, resulting in the probability
  distribution concentrated close to $0$ for low values of
  $r$. Similarly, we expect few cells to be infected by a very large
  number of IUs, accumulating the probability distribution close to
  $0$ for large $r$. Only at intermediate values of $r\approx N/M=10$
  we observe a Poisson-like distribution. (b) We perform a numerical
  study to show empirically that our analytical result in
  Eq.~\ref{EQ_PROB} matches the statistical frequency of virus-cell
  counts from a simulation of $N=100$ IUs being randomly assigned to
  $M=10$ cells. The square sum error between the simulated proportions
  and the analytical result was calculated with increasing numbers of
  iterations of the simulation. For iterations around $10^6$, our
  square sum error is on the order of $10^{-6}$, indicating strong
  agreement between our model and simulation.}
\label{FIG_ANALYTIC}
\end{center}
\end{figure}

We assume each viral particle in solution acts independently of others
and that host cell infection attempts are random events. At high
ratios $N_0/M$ of particles to cells, a quantity referred to as the
``multiplicity of infection'' (MOI), it becomes increasingly probable
for more than one IU to infect the same host cell. We define $M_0$
as the count of cells not infected by any IU, $M_1$ as the count of
cells infected by exactly one IU, up to $M_N$, the number of cells
infected by all $N$ IUs. The statistical multiplicity of infection
(SMOI) is defined as the ensemble of cell counts $\{M_0, M_1,\cdots,
M_N\}$. Note that two constraints must hold: $\sum_{r=0}^N M_r = M$ to
account for all infected and un-infected cells, and $\sum_{r=0}^N r
M_r = N$ for conservation of the total number of IUs. If we assume all
$M$ cells are of identical size and volume, they carry equal
probability of being infected by a particular virus. Thus,
  evaluating the probability distribution that $M_r$ takes on the
  value $m_r$ reduces to the well-known occupancy problem of randomly
  placing balls into identical urns \cite{Roberts2005} and we derive
\begin{widetext}
\begin{equation}\label{EQ_PROB}
\mathrm{Pr}(M_r = m_r\vert M, N) = \sum_{j=m_r}^M 
\binom{j}{m_r}\binom{M}{j} \binom{N}{r,\cdots,r,(N-rj)}\frac{(-1)^{j-m_r} 
\left(M-j\right)^{N-rj}}{M^N},
\end{equation}
\end{widetext}
where the $r$ term is repeated $j$ times in the lower argument of the
multinomial coefficient.
The derivation of Eq.~\ref{EQ_PROB} is detailed in
Appendix A in the SI and an investigation into the
effects of inhomogeneous cell sizes is presented in
Appendix B. Furthermore, in Appendix A,
we derive the expected value and variance of $M_r$ as
\begin{equation}
\mathrm{E}\left[M_r\right]  =   M\binom{N}{r}  
\left(\frac{1}{M}\right)^r\left(1 - \frac{1}{M}\right)^{N-r},
\end{equation}
and
\begin{align}
\mathrm{Var}\left[M_r\right] = M\binom{N}{r}  
\left(\frac{1}{M}\right)^r\left(1 - \frac{1}{M}\right)^{N-r} \hfill
\nonumber \\
\hspace{3mm} + \frac{M(M-1)N! (M-2)^{N-2r}}{(r!)^2 (N-2r)! M^N} 
\nonumber \\
- \frac{M^2 (N!)^2 (M-1)^{2N-2r}}{(r!)^2\left[(N-r)!\right]^2 M^{2N}}.
\end{align}
Note that the variance is equal to the expected value with two
additional correction terms that cancel each other as $N$ and $M$
increase, indicating the probability distribution of $M_r$ is
Poisson-like for large $N$ and $M$. A plot of a representative
  probability distribution and a test of agreement between our
  analytical result and numerical simulation is provided in
  Fig.~\ref{FIG_ANALYTIC}.

We also derive the joint probability
$\mathrm{Pr}(M_0=m_0,\cdots,M_N=m_N\vert M, N)$ that the SMOI
$\{M_0,M_1,\cdots,M_N\}$ takes on the set of values
$\{m_0,m_1,\cdots,m_N\}$ as
\begin{widetext}
\begin{eqnarray}\label{EQ_JOINT_PROB}
\mathrm{Pr}(M_0 = m_0,\cdots,M_N=m_N\vert M, N) &=& 
\frac{1}{M^N}\binom{M}{m_0,m_1,\cdots,m_N}
\binom{N}{0,\cdots,0,1,\cdots,1,\cdots,N,\cdots,N}\nonumber\\
&=& \frac{M!N!}{M^N} \prod_{r = 0}^N \frac{1}{m_r!\left(r!\right)^{m_r}}.
\end{eqnarray} 
\end{widetext}
The first and second multinomial expressions enumerate the degeneracy
of how the $M$ identical cells are distributed across the
configuration $\{m_0,\cdots,m_N\}$ and how the $N$ identical IUs are
chosen for those cells respectively. Although the second expression in
Eq.~\ref{EQ_JOINT_PROB} is more succinct, it must be explicitly
conditioned on the constraints $\sum_{r=0}^N m_r = M$ and
$\sum_{r=0}^N rm_r = N$.

\begin{figure*}[t!]
\begin{center}
\includegraphics[width=6.5in]{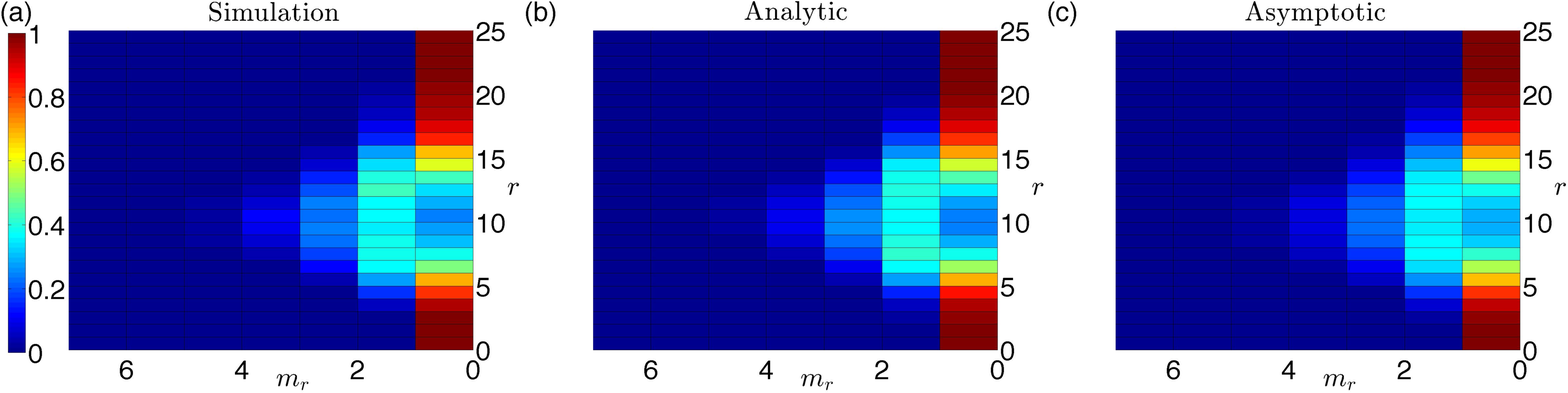}
\caption{Heat maps of the probability distribution
  $\mathrm{Pr}\left(M_r = m_r \vert M,N \right)$ of finding $m_r$
  cells that have been infected by exactly $r$ IUs given a total
  number of viruses $N=100$ and $M=10$ cells. (a) The statistical
  frequency of virus-cell counts after simulating IUs randomly
  distributing to the $M$ cells, averaged over 1000 iterations. (b) The
  analytical result obtained from Eq.~\ref{EQ_PROB}. (c) The
  asymptotic approximation with $M=10$ and $\mu=\frac{N}{M}=10$, using
  the expression in Eq.~\ref{EQ_PROB_APPROX}. There is close agreement
  between the simulated and analytical results. The relatively low
  values of $M$ and $N$ makes the asymptotic formula in
  Eq.~\ref{EQ_PROB_APPROX} inappropriate for this parameter regime,
  explaining the discrepancy between the asymptotic result and the
  exact analytical result. However, it is noteworthy how qualitatively
  small that deviation is, which will continue to vanish as $M$ and
  $N$ increase in value.}
\label{FIG_HEAT}
\end{center}
\end{figure*}

The expressions in Eqs.~\ref{EQ_PROB} and \ref{EQ_JOINT_PROB} provide
an exact discrete description of the stochasticity of the MOI, but are
computationally expensive to evaluate for large values of $N$ and
$M$. In a typical virology experiment, the number of viral particles
$N_0$ and host cells $M$ are large enough for certain asymptotic
methods to be applicable. Furthermore, for intermediate values of $Q$,
and based on Eq.~\ref{EQ_N_BINOM_PROB}, the expected number of IUs $N$
would be similarly large. We can thus take the mathematical limit $N,
M \to \infty$ while keeping the ratio $\mu=\frac{N}{M}$ fixed and
approximate Eq.~\ref{EQ_PROB} as:
\begin{equation}\label{EQ_PROB_APPROX}
\mathrm{Pr}(M_r = m_r \vert M, N) \approx \frac{1}{m_r!} \left[\frac{M \mu^r e^{-\mu}}{r!}\right]^{m_r}\!\!\!\exp\left[-\frac{M \mu^r e^{-\mu}}{r!}\right].
\end{equation}
Eq.~\ref{EQ_PROB_APPROX} implies that $M_r$ is Poisson-distributed with mean and variance
\begin{equation}
\label{EQ_POISSON_EXP}
\mathrm{E}[M_r] = \mathrm{Var}[M_r] \approx \frac{M \mu^r e^{-\mu}}{r!}.
\end{equation}
A mathematical justification of Eq.~\ref{EQ_PROB_APPROX} is given in
Appendix A and comparisons of Eq.~\ref{EQ_PROB_APPROX}
and the analytical result in Eq.~\ref{EQ_PROB} to simulations are
shown in Fig.~\ref{FIG_HEAT}.
Under the same large $M, N$ limit and using Eq.~\ref{EQ_PROB_APPROX},
we show in Appendix A
\begin{equation}
\begin{array}{l}
\label{EQ_JOINT_PROB_APPROX}
\mathrm{Pr}(M_0 = m_0,\cdots,M_N=m_N \vert M, N) \\[12pt]
\: \hspace{2cm}\approx \prod_{r=0}^N \mathrm{Pr}(M_r = m_r \vert M, N),
\end{array}
\end{equation}
which implies that as $M,N \to \infty$, the random variables
$M_0$,$\cdots$, $M_N$ are independently distributed. In the next
section, we will apply results of our probability model of SMOI to the
case of a repeatedly diluted solution of virus particles, a procedure 
used in many VQAs.

%-----------------------------------------------------------------------

\subsection{Serial Dilution}\label{SEC_SERIAL}
\begin{figure}[t!]
\begin{center}
\includegraphics[width=3.55in]{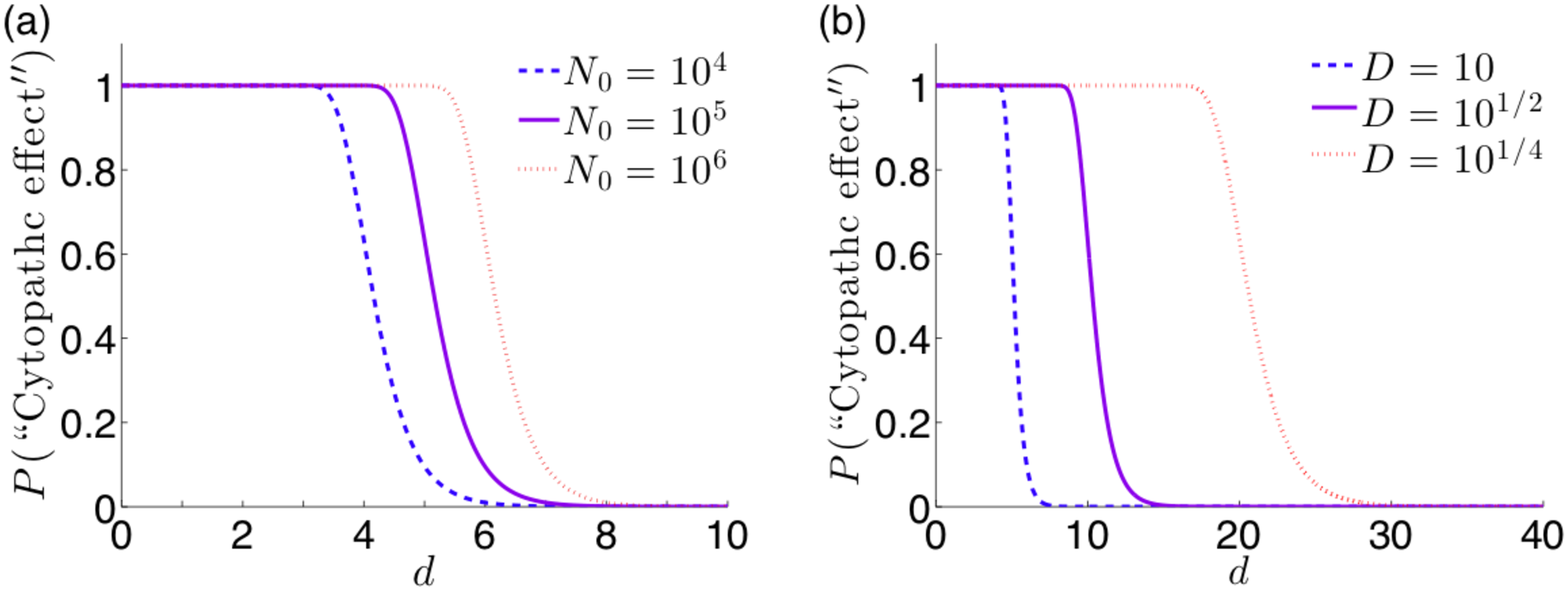}
\caption{The probability of observing a cytopathic effect (CE) given
  in Eq.~\ref{EQ_PROB_CYT} as a function of the dilution number $d$
  and with $Q = 1$. (a) For $D=10$, as the initial particle count
  $N_0$ increases, the critical dilution moves toward higher $d$. (b)
  Common dilution factors include logarithmic dilution ($D=10$),
  half-logarithmic dilution ($D=10^{1/2}$), and quarter-logarithmic
  dilution ($D=10^{1/4}$). Logarithmic dilution requires a lower
  number of dilutions to cause the characteristic decrease in
  probability, requiring less individual assays to
  perform. Quarter-logarithmic dilution, though requiring more
  dilutions, has a slower transition from high to low probability
  across $d$, making the assay less sensitive to experimental error or
  noise. The plot above can be used to quantify the tradeoffs between
  the choices of $D$.}
\label{FIG_CYTOPATH}
\end{center}
\end{figure}

Low viral particle concentrations in assays are typically obtained via
serial dilution processes in order to increase the sensitivity to
individual viral infections
\cite{Platt1,Kropinski2009,Johnson1990}. The initial viral stock
containing $N_0$ particles is diluted by a fixed factor of $D$ and the
process is repeated $d_{\mathrm{max}}$ times. At each dilution number
$d$, an assay can be performed to determine if the concentration of
virus particles in the diluted solution is sufficient to generate a
qualitative signal of infection, known as a ``cytopathic effect''
(CE). For example, the diluted stock can be administered \textit{in
  vivo} to a model organism such as a mouse. The mouse's death would
indicate that at least one lethal unit of the virus was present at
that dilution level. Alternatively, an \textit{in vitro} assay can be
carried out to measure a signal that, for example, quantifies the
exact number of plated cells that were successfully infected. To model
these assays, we first define $M^*$ as the number of host cells
infected by at least one IU and that are capable of producing new
viruses. In Appendix A we derive the discrete
probability density function for finding $M^*=m$ infected cells at a
given dilution number $d$ and find
\begin{equation}
\begin{array}{l}
\label{EQ_H_PROB_DIST}
\mathrm{Pr}\left(M^* = m\right) =  \binom{M}{m} 
\left[1 - \exp\left(-\frac{N_0}{QMD^d}\right)\right]^m \\[12pt]
\:\hspace{3cm}\times \exp\left(-\frac{N_0}{QMD^d}\right)^{M-m}.
\end{array}
\end{equation}
Eq.~\ref{EQ_H_PROB_DIST} shows that the number of infected cells $M^*$
is binomially distributed with expected value
\begin{equation}\label{EQ_H_EXPECTATION}
\mathrm{E}[M^*] = M\left[1 - \exp\left(-\frac{N_0}{QMD^d}\right)\right],
\end{equation}
and variance
\begin{equation}
\mathrm{Var}[M^*] = M\left[1 - \exp\left(-\frac{N_0}{QMD^d}\right)\right]
\exp\left(-\frac{N_0}{QMD^d}\right).
\end{equation}
We can define the probability of observing a CE at dilution number $d$
as the probability of finding one or more infected cells:
\begin{eqnarray}\label{EQ_PROB_CYT}
\mathrm{Pr}(\text{``Cytopathic effect''}) 
&\equiv& \sum_{m = 1}^M \mathrm{Pr}(M^* = m) \nonumber\\
&=& 1-\exp\left(-\frac{N_0}{Q D^d}\right).
\end{eqnarray}
The definition we use in Eq.~\ref{EQ_PROB_CYT} assumes an \textit{in
  vitro} assay that can exhibit a cytopathic signal after a single
cell infection or more. For \textit{in vivo} assays, the probability
that $m$ infected cells are sufficient for a CE will depend on many
complex physiological factors such as immune pressure, in-host viral
evolution, and virion burst size \cite{Gilchrist2004}. A plot of how
the initial particle count $N_0$ and dilution factor $D$ effect the
characteristic functional form of Eq.~\ref{EQ_PROB_CYT} are shown in
Fig.~\ref{FIG_CYTOPATH}. Although both Eqs.~\ref{EQ_H_PROB_DIST} and
\ref{EQ_PROB_CYT} assume each IU contains all viral genes required for
in-host replication, an extended probability model that factors in
genetic mutation and degradation is provided in Appendix C.
Furthermore, for the case of retroviruses, infectious processes inside
the host cytoplasm may be suppressed by previous infections, known as
viral interference, and is explored in
Appendix D. In Section~\ref{SEC_PLAQUE}, we will
use Eq.~\ref{EQ_H_PROB_DIST} to analyze the plaque
assay. Eq.~\ref{EQ_PROB_CYT} will be used for ``binary'' assays that
are only concerned with the presence or absence of a CE such as the
endpoint dilution assay, which we will explore in
Section~\ref{SEC_ENDPOINT}.

%%%%%%%%%%%%%%%%%%%%%%%%%%%%%%%%%%%%%%%%%%%%%%%%%%%%%%%
\section{Results and Discussion}
\subsection{Plaque Assay}\label{SEC_PLAQUE}
The plaque assay is an example of a virus quantification assay (VQA)
where the objective is to infer the total number of viruses $N_0$
present in a solution assuming the PFU to particle ratio $Q$ has been
independently measured and estimated
\cite{Kropinski2009,Johnson1990,Sloutskin2014}. After $d$ serial
dilutions, the viral stock is added to a monolayer of $M$ cells and a
layer of agar gel is added to the well to inhibit the diffusion of
virus particles in the plate. If a virus successfully infects a host
cell, the agar will limit the range of new infections to the most
adjacent cells. Viral infection thus spreads out radially from the
initial nucleation infection and forms a visible discoloration in the
plate called a ``plaque.'' For high particle concentrations, the
number of plaques formed may be large enough to cover the entire plate
surface. After a sufficient critical dilution number $d_c$ however,
the number of plaques formed are low enough to be visibly distinct and
countable. For each dilution number $d$, the assay can be performed
for $T$ number of trials. The `signal' data arising from the plaque
assay $P_{d,t}$ is defined as the number of visible plaques counted,
where $t=1,\cdots,T$ is the trial number. The standard method of
obtaining an estimate $\hat{N}_0$ of the true particle count $N_0$ is
to apply the sample mean of the data $P_{d_c,t}$ at the critical
dilution level $d_c$ to the formula
\begin{equation}\label{EQ_OLD_PLAQUE}
\hat{N}_0 = D^{d_c}\left(\frac{1}{T} \sum_{t = 1}^T P_{d_c,t}\right),
\end{equation}
which posits that the average number of plaques is directly
proportional to the particle count $N_0$. Eq.~\ref{EQ_OLD_PLAQUE}
assumes that each infected cell corresponds to one IU, which is not
necessarily true in the context of SMOI. Furthermore, although data
corresponding to dilution numbers $d < d_c$ are unusable, data for
$d>d_c$ corresponding to countable plaques are not used at all in
Eq.~\ref{EQ_OLD_PLAQUE}.

\begin{figure}[t!]
\begin{center}
\includegraphics[width=3.5in]{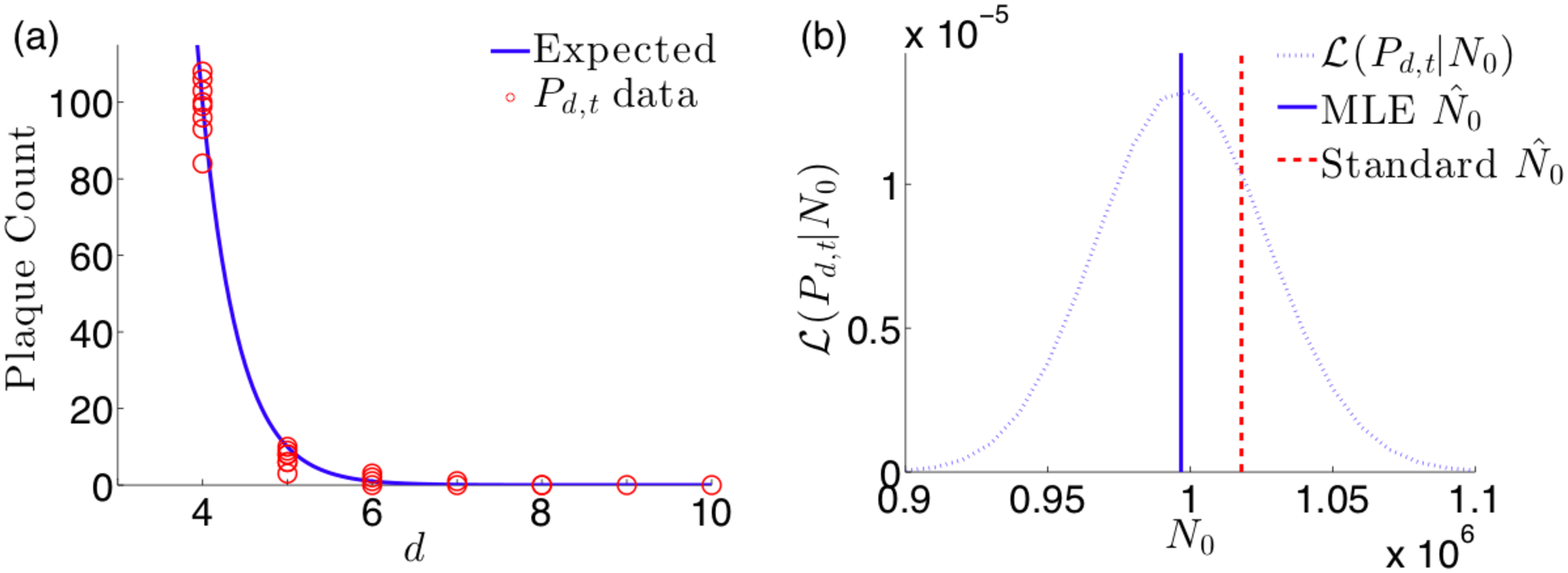}
\caption{Results of plaque assay simulation for parameters $N_0=10^6$,
  $M=10^5$, $Q=1$, $D=10$, $d_{\mathrm{max}}=10$, and $T=10$. (a) The
  scatter plot of simulated data $P_{d,t}$ (circles) and the expected
  value of plaque counts as given by Eq.~\ref{EQ_H_EXPECTATION} show
  close agreement. (b) The likelihood function
  $\mathcal{L}(P_{d,t}\vert N_0)$ with respect to $N_0$ using the same
  simulated data. The MLE obtained by iteratively solving
  Eq.~\ref{EQ_PLAQUE_NEWTON} is $\hat{N}_0 = 9.97\times 10^5$ and is
  relatively closer to the true value of $N_0$ than the estimate
  calculated from the standard method in Eq.~\ref{EQ_OLD_PLAQUE}
  $\hat{N}_0 = 1.02\times 10^6$.}
\label{FIG_SIM_PLAQUE}
\end{center}
\end{figure}

In order to improve on Eq.~\ref{EQ_OLD_PLAQUE} by using the entire set
of plaque counts $P_{d,t}$ for our estimate of $N_0$, we propose a
maximum likelihood estimation (MLE) scheme. Using the mathematical
models derived above, we can construct an expression
$\mathcal{L}(P_{d,t}\vert N_0)$ of the probability that the observed
data $P_{d,t}$ can be generated assuming a particular value for $N_0$,
known as a likelihood function. A value for $N_0$ that maximizes
$\mathcal{L}(P_{d,t}\vert N_0)$ corresponds to the most probable
estimate $\hat{N}_0$ that could have generated the data. As each
nucleation of a plaque corresponds to a distinct infected cell (and
assuming that overlapping lesions of necrotic cells are still
discernible as distinct plaques), we can equate $P_{d,t}$ to the total
number of successfully infected cells $M^*$. We will ignore the
dynamics of coinfection and viral interference. Using
Eq.~\ref{EQ_H_PROB_DIST}, we propose the following likelihood function
of the data given $N_0$:
\begin{widetext}
\begin{equation} \label{EQ_PLAQUE_LIKELIHOOD}
\mathcal{L}(P_{d,t}\vert N_0) = \prod_{d = d_c}^{d_{\mathrm{max}}} 
\prod_{t = 1}^T \binom{M}{P_{d,t}} 
\left[1 - \exp\left(-\frac{N_0}{QMD^d}\right)\right]^{P_{d,t}} 
\exp\left(-\frac{N_0}{QMD^d}\right)^{M-P_{d,t}}.
\end{equation}
\end{widetext}
To obtain the MLE $\hat{N}_0$, we take the derivative of the natural
log of Eq.~\ref{EQ_PLAQUE_LIKELIHOOD} with respect to $N_0$ and set
the result to zero to obtain
\begin{equation} \label{EQ_PLAQUE_NEWTON}
0 = \sum_{d = d_c}^{d_{\mathrm{max}}} \sum_{t = 1}^T 
\frac{M \exp\left(-\frac{\hat{N}_0}{QMD^d}\right) - 
M + P_{d,t}}{QMD^d\left[1 - \exp\left(-\frac{\hat{N}_0}{QMD^d}\right)\right]}.
\end{equation}
We can solve Eq.~\ref{EQ_PLAQUE_NEWTON} for $\hat{N}_0$ using
numerical methods such as Newton-Raphson \cite{Lange2013}, an
iterative scheme that approaches the solution of an equation
asymptotically starting from an initial guess
$\hat{N}_0^{\mathrm{init}}$. To increase the stability of convergence
to the solution, we choose $\hat{N}_0^{\mathrm{init}}$ by equating the
sample average of plaque counts $\frac{1}{T}\sum P_{d_c,t}$ with the
expected number of infected cells $\mathrm{E}[M^*]$ in
Eq.~\ref{EQ_H_EXPECTATION} at the critical dilution $d_c$ to derive
\begin{equation}
\hat{N}_0^{\mathrm{init}} = -QMD^{d_c} 
\ln\left[1 - \frac{1}{M}\left(\frac{1}{T}\sum_{t = 1}^T P_{d_c, t}\right)\right].
\end{equation}
\begin{figure}[t!]
\includegraphics[width=2.6in]{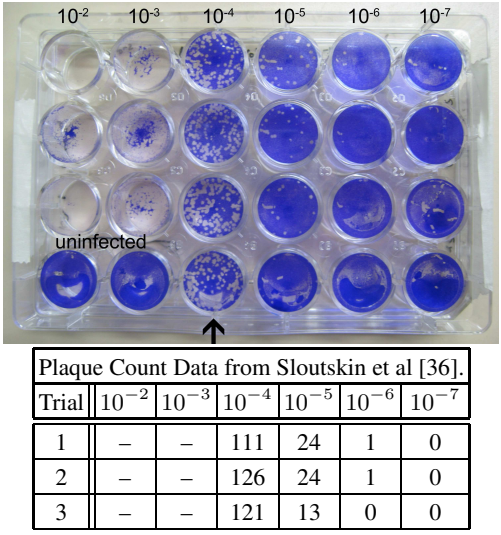}
\\
%\: \hspace{3in}
\begin{tabular}[b]{|c||c|c|c|c|c|c|}
 \hline
 \multicolumn{7}{|c|}{Plaque Count Data from Sloutskin et al \cite{Sloutskin2014}.} \\
 \hline
 Trial & $10^{-2}$ & $10^{-3}$ & $10^{-4}$ & $10^{-5}$ & $10^{-6}$ & $10^{-7}$ \\
 \hline
 \hline
 1   & --    & -- & 111 & 24 & 1 & 0\\
 \hline
 2   & --    & -- & 126 & 24 & 1 & 0\\
 \hline
 3   & --    & -- & 121 & 13 & 0 & 0\\
 \hline
\end{tabular}
\caption{An example of raw plaque count data taken from Sloutskin et
  al. \cite{Sloutskin2014}. A viral solution was assayed in a plate of
  $M=3\times 10^5$ cells at dilution numbers $d=2,3,4,5,6,$ and $7$ at a
  dilution factor of $D=10$. The particle to PFU ratio is assumed to
  be $Q=1$. For $T=3$ separate trials, the number of plaques were
  counted at each dilution level. The bottom row of plates used as a
  control is ignored. For dilution numbers $d=2$ and $3$, the entire
  plate of cells show cytotoxicity so that the numbers of plaques were
  undiscernable and, thus, the countable data starts at $d_c = 4$. For
  the old method featured in Eq.~\ref{EQ_OLD_PLAQUE}, the estimate for
  $N_0$ is $\hat{N}_0 = 1.19\times 10^6$ and for the MLE derived from
  Eq.~\ref{EQ_PLAQUE_NEWTON}, $\hat{N}_0 = 1.26\times 10^6$. This results
  in a relative difference of 5.5\%. Furthermore, when applying these
  parameters and $\hat{N}_0$ estimate to Eq.~\ref{EQ_PLAQUE_VARIANCE},
  we observe a 10.7\% decrease in the estimate variation using the MLE
  technique.}
\label{FIG_PLAQUE}
\end{figure}
An example of raw plaque count data and the resulting estimates for
$N_0$ are given in Fig.~\ref{FIG_PLAQUE}. In order to quantify the
relative improvement of the MLE of $N_0$ over the standard method in
Eq.~\ref{EQ_OLD_PLAQUE}, we simulate plaque assay data assuming a
fixed, known $N_0$ value. In our simulation, we use the models
established in Section~\ref{SEC_PROBABILITY} to sample the $N_0$
particles according to Eq.~(S9) in
Appendix A to account for serial dilution and sample
again the resulting particles according to Eq.~\ref{EQ_N_BINOM_PROB}
to obtain the number of IUs $N$. The IUs are distributed randomly to
the $M$ cells with equal probability and the resulting number of
infected cells $M^*$ is recorded. Since plates of cells with too many
infections render the number of plaques uncountable, a ``countable
plaque threshold'' renders the data unusable when the number of
infected cells exceed the threshold.  Thus, the resulting plaque data
$P_{d,t}$ for a given dilution $d$ and trial $t$ is assigned the
number of simulated infected cells if the latter is less than the
given threshold. A scatter plot of the data $P_{d,t}$ of one such
simulation is shown in Fig.~\ref{FIG_SIM_PLAQUE}a and the
corresponding likelihood function from Eq.~\ref{EQ_PLAQUE_LIKELIHOOD}
is plotted in Fig.~\ref{FIG_SIM_PLAQUE}b. Because the MLE method
utilizes a full probabilistic model of the plaque count distribution
instead of relying only on the expected value at the single critical
dilution $d_c$, it produces an estimate consistently closer to the
original $N_0$ that generated the data. To better quantify this
property, in Appendix E we derive an asymptotic
approximation of the variance of $\hat{N}_0$ as
\begin{equation}\label{EQ_PLAQUE_VARIANCE}
\mathrm{Var}\left[\hat{N}_0\right] \approx \left[\sum_{d =
    d_{\mathrm{c}}}^{d_{\mathrm{max}}} \frac{T \exp\left(-\frac{N_0}{Q
      M D^d}\right)}{Q^2 M D^{2d}\left[1 - \exp\left(-\frac{N_0}{Q M
        D^d}\right)\right]}\right]^{-1}.
\end{equation}
The variance is an explicit function of $Q$, which is assumed to be
\textit{a priori} known. If there is uncertainty in the value of $Q$,
Eq.~\ref{EQ_PLAQUE_VARIANCE} can quantify how sensitive the
distribution of $\hat{N}_0$ is to variation in $Q$, as shown in
Fig.~\ref{FIG_FISHER}a. We can see that for small assumed $Q$, such as
in poliovirus \cite{Klasse2015}, error in this measurement can cause a
large relative change in the accuracy of $\hat{N}_0$. This type of
sensitivity analysis on estimation variance can be done with any
experimental parameter included in the likelihood function in
Eq.~\ref{EQ_PLAQUE_LIKELIHOOD}. Furthermore, for directly controllable
parameters, such as the serial dilution factor $D$,
Eq.~\ref{EQ_PLAQUE_VARIANCE} can provide insight into optimizing the
assay protocol, as shown in Fig.~\ref{FIG_FISHER}b. Although it is
evident that small $D$ would increase the accuracy of the $\hat{N}_0$
estimate, doing so requires more serial dilutions which increases the
time and expense of the assay. Thus, our sensitivity analysis provides
a quantitative method for making experimental design choices between
minimizing uncertainty versus the cost of an assay protocol. Lastly,
if we compute the variance of the standard method in
Eq.~\ref{EQ_OLD_PLAQUE} due to the known variance in the data
$P_{d,t}$, and compare with Eq.~\ref{EQ_PLAQUE_VARIANCE}, we find,
when using realistic parameter values from Fig.~\ref{FIG_PLAQUE}, the
standard method results in a $10.7\%$ higher variance than that of our
method.  Although the significance of the relative increase in
precision of estimating $N_0$ found using our method is highly
dependent on the context of the experimental study for which the assay
was performed, similar sensitivity analysis can be used to determine
such tolerances.
\begin{figure}[t!]
\begin{center}
\includegraphics[width=3.5in]{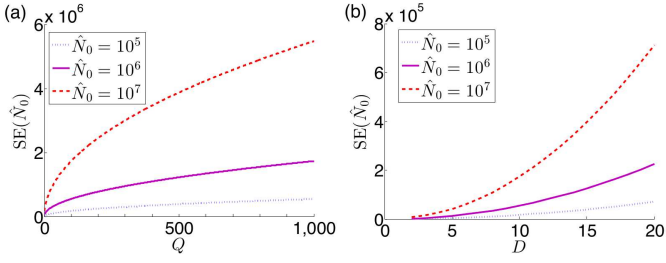}
\caption{Approximations of the standard deviation
    $\sigma_{\hat{N}_0} =\mathrm{Var}[\hat{N}_0]^{1/2}$ of maximum
    likelihood estimates for the plaque assay using
    Eq.~\ref{EQ_PLAQUE_VARIANCE} and parameters $\hat{N}_0=10^5$,
    $10^6$, and $10^7$, $M=3\times 10^5$, $d_{\mathrm{c}}=4$,
    $d_{\mathrm{max}}=7$, and $T=3$, corresponding to the assay
    displayed in Figure~\ref{FIG_PLAQUE}. (a) For $D=10$, the standard
    deviation increases proportional to the square root of $Q$. (b)
    For $Q=1$, we can see a low dilution factor $D$ will increase the
    accuracy of the estimate $\hat{N}_0$.}
\label{FIG_FISHER}
\end{center}
\end{figure}
%

%%%%%%%%%%%%%%%%%%%%%%%%%%%%%%%%%%%%%%%%%%%%%%%%%%%%%%%
\subsection{Endpoint Dilution Assay}\label{SEC_ENDPOINT}
\begin{figure}[t!]
\begin{center}
\includegraphics[width=3.4in]{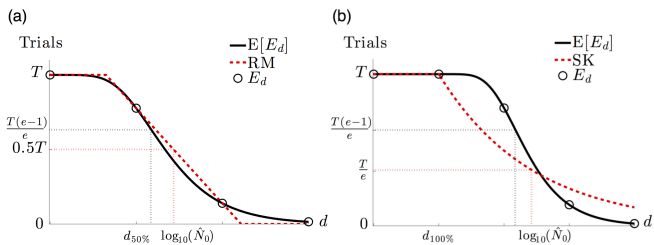}
\caption{An illustration of the consistent 
overestimation of the Reed and Muench (RM) and Spearman-Karber (SK)
methods using the expected curve $\mathrm{E}[E_d]$ of CEs given $T$
trials as a function of the dilution number $d$ derived from
Eq.~\ref{EQ_PROB_CYT}. (a) The RM method approximates the steepest
decent of the expectation curve with a line connecting the two data
points $E_{d_{50\%}}\leq 0.5T <E_{d_{50\%}+1}$. Because of the
relative convexity of the expected curve, the linear approximation
consistently rests above the curve and results in an overestimate of
$\log_{10}(\hat{N}_0)$. (b) From the last dilution $d_{100\%}$ such
that all trials exhibit a CE, the SK method assumes an exponential
decay of the expectation. Obtaining the characteristic decay rate of
the exponential involves calculating the area under the curve, which
is done numerically using the data $E_d$. However, according to our
model, many of the expected values of $E_d$ exist above the
exponential, causing the numerical integration to overestimate the
area and, thus, decay too slowly. This gradual decrease in the
exponential curve results in a larger estimate of
$\log_{10}(\hat{N}_0)$.}
\label{FIG_RESIDUAL_ED}
\end{center}
\end{figure}

Another widely used assay for quantifying the initial viral particle
count $N_0$ is the endpoint dilution or endpoint titration assay
\cite{Ramakrishnan2016,Neumann2005,Johnson1990}. It is often used in
place of the plaque assay as it can be more rapidly performed and is
useful for viral strains that are unable to form plaques. Here, serial
dilutions at a factor of $D$ are employed and at every dilution number
$d$, an assay is performed $T$ times to test for a successful CE. The
number $E_d$ of observed CEs among the $T$ trials at a given dilution
number $d$ is recorded as the signal. For low dilution, we expect many
cells to be infected and the probability of observing a CE, as shown
in Eq.~\ref{EQ_PROB_CYT}, is close to $1$. If every trial of the assay
is likely to display a CE, then $E_d$ is expected to be close to $T$.
However, at high dilution, the probability in Eq.~\ref{EQ_PROB_CYT}
rapidly decreases to $0$, as shown in Fig.~\ref{FIG_CYTOPATH}, and
$E_d$ will be similarly small.  For a large initial stock of viral
particles $N_0$, a larger dilution number $d$ is needed to ensure the
dramatic change in probability in Eq.~\ref{EQ_PROB_CYT}.  Thus, the
critical dilution at which $E_d$ most rapidly decreases from $T$ can
be used to estimate the particle count $N_0$. This occurs at the point
of inflection when $d=\log_D(N_0Q^{-1})$ and corresponds to when the
expected number of successful trials $\mathrm{E}[E_d] = T(1-e^{-1})$,
as shown in Fig.~\ref{FIG_RESIDUAL_ED}.

One commonly used way to estimate $N_{0}$ is the Reed and Muench (RM)
method that utilizes the two dilution numbers that capture the
greatest change in the data $E_{d}$ \cite{Reed1938}. We first define a
critical dilution number $d_{50\%}$ to be the largest dilution such
that at least $50\%$ of the trials exhibit a CE. The estimate
$\hat{N}_0$ for the particle count $N_0$ is given by
\begin{equation}
\log_{10}(\hat{N}_0) = d_{50\%} + \frac{E_{d_{50\%}} - 0.5T}{E_{d_{50\%}} - E_{d_{50\%}+1}}.
\end{equation}
The RM method effectively attempts to approximate the 
steepest descent of the CE probability given in Eq.~\ref{EQ_PROB_CYT}
with a line connecting the assay data at dilutions $d_{50\%}$ and
$d_{50\%+1}$, as displayed in
Fig.~\ref{FIG_RESIDUAL_ED}a. Unfortunately, this line always rests
above the actual expectation curve of $E_d$, so any estimate
$\hat{N}_0$ obtained from this method will overestimate the true
$N_0$. Another commonly used estimation scheme is the Spearman-Karber
(SK) method which uses the critical dilution number $d_{100\%}$, the
largest dilution such that $100\%$ of trials exhibit a cytopathic
effect \cite{Hamilton1977,Ramakrishnan2016}. The SK estimate
$\hat{N}_0$ is given by
\begin{equation}\label{EQ_SK}
\log_{10}(\hat{N}_0) = d_{100\%} - \frac{1}{2}\log_{10}(D) + 
\log_{10}(D)\sum_{d = d_{100\%}}^{d_{\mathrm{max}}}\!\!\!\frac{E_d}{T}.
\end{equation}
In this method, the downward slope for the expectation 
of $E_d$ is assumed to follow a decaying exponential starting at
dilution $d_{100\%}$, as shown in Fig.~\ref{FIG_RESIDUAL_ED}b. The
intention is to find the dilution at which $Te^{-1}$ CEs are expected
by calculating the area under the exponential curve, given by the
summation term in Eq.~\ref{EQ_SK}. However, the actual values of $E_d$
will follow the expected curve from our model, leading to an
overestimate of the area and, by extension, a larger value for
$\hat{N}_0$. Both standard methods were derived from the heuristic
observation that $E_d$ exhibits sigmoidal behavior as a function of
the dilution number $d$, but an underlying probabilistic model was
missing, resulting in consistent overestimation of the true $N_0$.
Furthermore, neither method uses the ``particle to PFU ratio'' $Q$,
accounts for the stochasticity of serial diluting viral samples,
considers the dynamics of SMOI, or employs the entire set of data
$E_d$.

%Another commonly used estimation scheme is the Spearman-Karber (SK)
%method which uses the critical dilution number $d_{100\%}$, the
%largest dilution such that $100\%$ of trials exhibit a cytopathic
%effect \cite{Hamilton1977,Ramakrishnan2016}. The SK estimate
%$\hat{N}_0$ is given by
%
%\begin{equation}
%\log_{10}(\hat{N}_0) = d_{100\%} - \frac{1}{2}\log_{10}(D) 
%+ \log_{10}(D)\sum_{d = d_{100\%}}^{d_{\mathrm{max}}}\frac{E_d}{T}.
%\end{equation}
% 
%Both methods are derived from the heuristic observation that 
%$E_d$ exhibits sigmoidal behavior as a function of the dilution
%number $d$, but an underlying probabilistic model is missing.
%Neither method uses the ``particle to PFU ratio'' $Q$, accounts for the
%stochasticity of serial diluting viral samples, or considers the
%dynamics of SMOI. Furthermore, both methods fail to employ the entire
%set of data $E_d$.

\begin{figure}[t!]
\begin{center}
\includegraphics[width=3.5in]{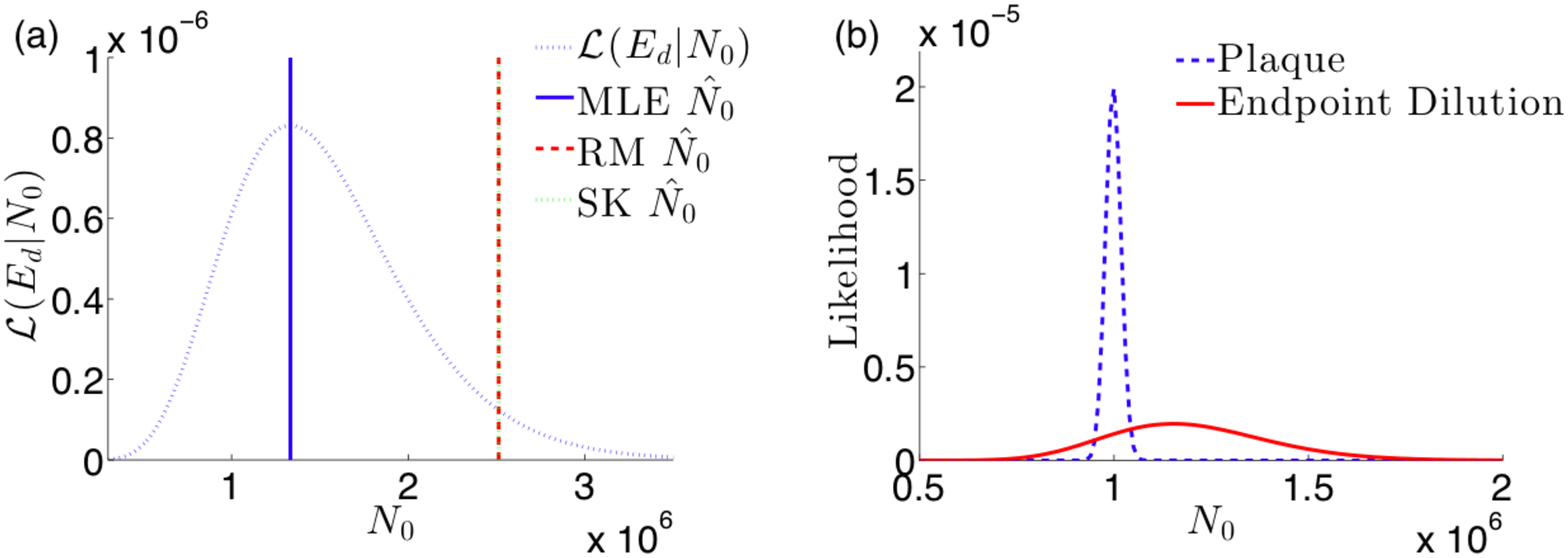}
\caption{(a) The likelihood function $\mathcal{L}(E_d \vert N_0)$ in
  Eq.~\ref{EQ_EDA_LIKELIHOOD} for the endpoint dilution assay and the
  corresponding maximum likelihood, Reed and Muench, and
  Spearman-Karber estimates given simulated data generated with
  $N_0=10^6$, $Q=1$, $D=10$, and $d_{\mathrm{max}} = 10$. The
  estimates for maximum likelihood ($\hat{N}_0=1.33\times 10^6$), RM
  ($\hat{N}_0=2.51\times 10^6$), and SK ($\hat{N}_0=2.51\times 10^6$) all
  overestimate $N_0$, but the smaller relative error of the MLE is an
  improvement on the errors of the existing two methods. (b) The
  likelihood functions $\mathcal{L}(P_{d,t} \vert N_0)$ and
  $\mathcal{L}(E_d \vert N_0)$ for the plaque and endpoint dilution
  assays respectively given simulated data. The data was generated
  with parameters $N_0=10^6$, $M=10^5$, $Q=1$, $D=10^{1/4}$,
  $d_{\mathrm{max}} = 30$, and a ``countable plaque threshold'' of
  $150$. The plaque assay likelihood is concentrated close to the true
  $N_0$ value while the endpoint dilution likelihood is far more
  spread out and overestimates $N_0$. This direct quantitative
    comparison can inform an experimentalist when choosing between the
    two methods.}
\label{FIG_SIM_ENDPOINT}
\end{center}
\end{figure}

We present an alternative way to infer $N_0$ using
Eq.~\ref{EQ_PROB_CYT} to establish a maximum likelihood estimation
scheme. We restrict ourselves to \textit{in vitro} assays in which a
single infected cell is sufficient to display a CE. Then each
cytopathic count is binomially distributed with parameters $T$ and the
probability given in Eq.~\ref{EQ_PROB_CYT}. Thus, for a set of data
$\{E_1,E_2,\cdots,E_{d_{\mathrm{max}}}\}$, we propose the likelihood
function
\begin{equation}\label{EQ_EDA_LIKELIHOOD}
\mathcal{L}(E_d \vert N_0) = \prod_{d = 1}^{d_{\mathrm{max}}}\binom{T}{E_d}\left[
e^{{N_{0}\over Q D^{d}}}-1\right]^{E_{d}}e^{-{TN_{0}\over Q D^{d}}}.
%
%1-\exp\left(-\frac{N_0}{QD^d}\right)\right]^{E_d}
%\exp\left(-\frac{N_0}{QD^d}\right)^{T - E_d}.
\end{equation}
Eq.~\ref{EQ_EDA_LIKELIHOOD} is an expression of the probability of the
data $\{E_1,\cdots,E_{d_{\mathrm{max}}}\}$ given the current assumed
value of $N_0$. To obtain the best estimate $\hat{N}_0$ of $N_0$, we
maximize the likelihood function by taking the log and derivative of
$\mathcal{L}(E_d \vert N_0)$ with respect to $N_0$ and set it equal to
zero to obtain
\begin{equation}\label{EQ_EDA_LOGLIKEDER}
0=\sum_{d=1}^{d_{\mathrm{max}}}\frac{E_d - T + T\exp\left(-\frac{\hat{N}_0}{QD^d}\right)}{QD^d\left(1 - \exp\left(-\frac{\hat{N}_0}{QD^d}\right)\right)}.
\end{equation}
As with Eq.~\ref{EQ_PLAQUE_NEWTON}, solving
Eq.~\ref{EQ_EDA_LOGLIKEDER} for $\hat{N}_0$ requires a numerical
method such as Newton-Raphson. As an appropriate initial estimate for
$\hat{N}_0$, the formula
\begin{equation}\label{EQ_EDA_INITIAL}
\hat{N}_0^{\mathrm{init}} = -0.5 Q D^{d_c}\left[\ln\left(1 - \frac{E_{d_c}}{T}\right) + D\ln\left(1 - \frac{E_{d_c+1}}{T}\right)\right],
\end{equation}
can be used, where $d_c$ is the largest dilution number such that at
least half of the trials exhibit a cytopathic
effect. Eq.~\ref{EQ_EDA_INITIAL} is the average of the $N_0$ estimates
at dilutions $d_c$ and $d_{c+1}$ when setting the CE probability in
Eq.~\ref{EQ_PROB_CYT} to $1/2$. For a comparison of our MLE method
with the RM and SK methods, we simulate data similar to that described
in Section~\ref{SEC_PLAQUE}. Here we take the number of trials such
that the simulated count of infected cells is greater than zero as the
values of $E_d$ for a given dilution number $d$. We plot the
likelihood from Eq.~\ref{EQ_EDA_LIKELIHOOD} and compare the MLE of
$N_0$ with those derived by the RM and SK methods in
Fig.~\ref{FIG_SIM_ENDPOINT}a. While both RM and SK estimate very similar
values of $\hat{N}_0$, they both consistently over-estimate the
\textit{a priori} set $N_0$ relative to the MLE method. This
demonstrates the advantage of a probabilistic model for parameter
inference over heuristically determined formulas.

The expressions we derived in Eqs.~\ref{EQ_PLAQUE_LIKELIHOOD} and
\ref{EQ_EDA_LIKELIHOOD} applied to simulated data can also help
quantify tradeoffs in experimental design. As discussed above, there
exist viruses that cannot form plaques, restricting the options of
VQAs to endpoint dilution. However, for many cases, the choice between
using one assay over the other can be one of convenience. More
specifically, endpoint dilution assays can often be performed more
rapidly than plaque assays. Using the same simulated data for both
assays, we plot Eqs.~\ref{EQ_PLAQUE_LIKELIHOOD} and
\ref{EQ_EDA_LIKELIHOOD} together in Fig.~\ref{FIG_SIM_ENDPOINT}b. The
plots clearly show the superiority of the plaque assay for estimating
the viral stock number $N_0$ in respect to both how close the MLE
infers the true $N_0$ value and the amount of variance in that
estimate. While the amount of variability and error that is tolerable
for an experiment may be context-dependent, the plots in
Fig.~\ref{FIG_SIM_ENDPOINT}b provide a quantitative way to
differentiate between the two methods.

%\begin{table}[t!]
%\begin{tabular}{ |c||c|c|c|c|c|c|c|  }
% \hline
% \multicolumn{8}{|c|}{Endpoint Dilution Lethal Mouse Data \cite{Ramakrishnan2016}} \\
% \hline
% Dilution Factor ($D^{-d}$) & $10^{-1}$ & $10^{-2}$ & $10^{-3}$ & $10^{-4}$ & $10^{-5}$ & $10^{-6}$ & $10^{-7}$ \\
% \hline
% \hline
% Mice Inoculated ($T$)   & 10 & 10 & 10 & 10 & 10 & 10 & 10 \\
% \hline
% Mice Died ($S_d$)   & 10 & 10 & 10 & 10 & 10 & 6 & 1 \\
% \hline
%\end{tabular}
%\caption{Example data for an endpoint dilution assay involving mice inoculated with a virus dissolved solution \cite{Ramakrishnan2016}. A successful cytopathic effect is a lethal result for the mouse. At each dilution, $10$ mice are inoculated and the number of cytopathic events, $S_d$ is counted. The resulting estimate using the iterative method detailed in Eq.~\ref{EQ_EDA_LOGLIKEDER} is $N_0=\num{9.36e5}$. The rough, initial approximation detailed in Eq.~\ref{EQ_EDA_INITIAL} is $N_0 = \num{8.75e5}$. Using the Reed and Muench method \cite{Reed1938} results in $N_0=\num{1.58e6}$ while the Spearman-Karber method \cite{Hamilton1977} yields $N_0=\num{1.58e6}$. This indicates that the two established methods overshoot the estimate of $N_0$ by an enormous relative margin of $69\%$.}
%\label{TABLE_ENDPOINT}
%\end{table}

%%%%%%%%%%%%%%%%%%%%%%%%%%%%%%%%%%%%%%%%%%%%%%%%%%%%%%%
\subsection{Luciferase Reporter Assay}\label{SEC_LUCIFERASE}
The luciferase reporter assay is commonly used to measure the
infectivity of a viral strain. Here the ratio $\mu = N/M$ of total
infections over the number of plated cells is estimated by measuring
the transcription activity of viral proteins
\cite{Chikere201381,Johnston,Webb2016}. The reporter employs an
oxidative enzyme luciferase that facilitates a reaction when
introduced to the substrate luciferin, resulting in
bioluminescence. The protocol begins with attaching the luciferase
gene to the viral genome. The altered viral strain is cloned to a
total particle count $N_0$ which, in this case, is assumed to be fixed
and known. The solution of viruses is added to a plated monolayer of
$M$ host cells. An incubation time is allowed for transcription of
viral proteins and, incidentally, the luciferase enzyme. Subsequently
all cells are lysed to release all cytoplasmic contents into the
solution upon which luciferin is added. The oxidation of luciferin is
facilitated by the luciferase enzyme and the resulting bioluminescence
yields a measurable signal \cite{Montefiori2009}. The light intensity
is thus a measure of total transcription activity of the viral genome
in all infected cells and can be used as a proxy for the total number
of viruses $N$ that successfully infected host cells.

\begin{figure}[t!]
\begin{center}
\includegraphics[width=3.6in]{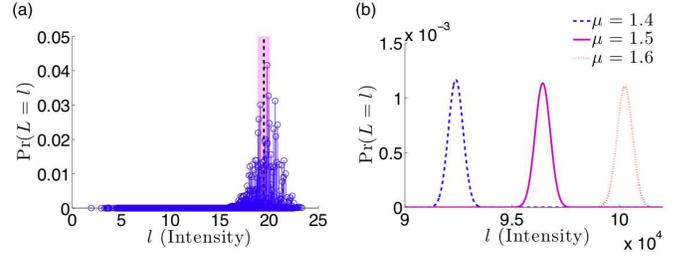}
\caption{Probability distributions of the luciferase assay
  fluorescence intensity $L$ from Eq.~\ref{EQ_LUC}. (a) A toy example
  of a discrete probability distribution of allowable fluorescence
  intensities for $N=30$ viruses infecting $M=20$ cells. Due to the
  $M^N$ finite number of allowable configurations of the SMOI, there
  are a corresponding finite number of intensities with specific
  probabilities determined by Eq.~\ref{EQ_JOINT_PROB} and represented
  by a unique circle. The parameters used for the reporter kinetics
  are $f_{\mathrm{max}} = 2$, $h=1$, $K=1$ and $L_0 = 1$. The mean
  intensity of the fluorescence signal is $\mathrm{E}[L] = 19.5$,
  represented by the vertical dotted line, and variance
  $\mathrm{Var}[L]=1.49$, represented by the shaded region. (b) The
  normally distributed approximation of fluorescence intensity using
  $M=10^5$, $f_{\mathrm{max}} = 2$, $h=1$, $K=1$ and $L_0 = 1$. The
  distributions are plotted for $\mu=1.4$, $1.5$, and $1.6$ by
  computing the expected values $\mathrm{E}[L] = 9.23\times 10^4$,
  $9.64\times 10^4$, and $10^5$ and the variances $\mathrm{Var}[L] =
  1.7\times 10^{5}$, $1.24\times 10^5$, and $1.3\times 10^5$
  respectively.}
\label{FIG_LUC}
\end{center}
\end{figure}

Although there is stochasticity in transcription factor binding and,
in the case of retroviruses, the number of integration sites on the
host DNA, we will assume that each successful virus infection
contributes one viral genome to be transcribed and each transcription
occurs at a constant rate proportional to the total number of
integrated viral genomes. Note that the limited number of
transcription factors, ribosomes, and other cell machinery necessary
to produce viral proteins and the luciferase reporter causes the
production rate to saturate as the number of infecting viruses $r$ per
cell increases. Thus, transcription activity saturates 
with increasing number of infections $r$. We can model this effect by
defining a monotonically increasing function $f(r)$ representing the
number of transcribed viral proteins when a cell is infected by $r$
viruses over the course of the assay. Thus, for a given SMOI
$\{M_0,\cdots,M_N\}$, we will model the intensity signal $L$ of the
total luciferase reporter luminescence with
\begin{equation}\label{EQ_LUC}
L=\sum_{r=0}^{N}L_0 f(r) M_r,
\end{equation}
where $L_0$ is the fluorescence intensity arising from a single
luciferase reporter present in the solution. Although $f(r)$ may take
on many functional forms, a commonly used model for transcription
factor kinetics is the Hill function \cite{Weiss1997} given by
\begin{equation}\label{EQ_HILL}
f(r) = \frac{f_{\mathrm{max}} r^h}{K + r^h},
\end{equation}
where $f_{\mathrm{max}}$ is the maximum transcription activity of
luciferase, $h$ is the Hill coefficient that effectively describes cooperative
binding of multiple transcription factors at a promoter region, and
$K$ is an effective dissociation constant relating the binding and
unbinding rates of transcription factor. The functional form of
Eq.~\ref{EQ_HILL} accounts for the limited transcription machinery
available for the multiple copies of viral genome present in the
cell. In Fig.~\ref{FIG_LUC}a we calculate the discrete probability
distribution $\mathrm{Pr}(L = \ell)$ by considering the cumulative weight
of every allowable configuration of $N$ viruses infecting $M$ cells
through Eq.~\ref{EQ_LUC}.

Since luciferase reporter assays typically involve large values of
initial virus count $N_0$ and cell count $M$, we can use the
asymptotic approximations in Eqs.~\ref{EQ_PROB_APPROX} and
\ref{EQ_POISSON_EXP} along with the Central Limit Theorem
\cite{Lange2003} to assume $L$ is normally distributed with expected
value
\begin{equation}\label{EQ_LUC_EXP}
\mathrm{E}[L] = L_0 f_{\mathrm{max}} Me^{-\mu}\sum_{r=0}^{N}\frac{r^h \mu^r}{(K + r^h)r!},
\end{equation}
and variance
\begin{equation}\label{EQ_LUC_VAR}
\mathrm{Var}[L] = L_0^2 f_{\mathrm{max}}^2 
Me^{-\mu}\sum_{r=0}^{N}\frac{r^{2h} \mu^r}{(K + r^h)^2 r!}.
\end{equation}
A visualization of the normal approximation of the probability
distribution of $L$ is shown in Fig.~\ref{FIG_LUC}b. Furthermore, with
Eqs.~\ref{EQ_LUC_EXP} and \ref{EQ_LUC_VAR}, we can derive the
likelihood function $\mathcal{L}(L_t^{\mathrm{data}} \vert \mu)$ of
the data $L_t^{\mathrm{data}}$, given $\mu$
\begin{equation}\label{EQ_LUC_LIKELIHOOD}
\mathcal{L}(L_t^{\mathrm{data}} \vert \mu) = \prod_{t=0}^T 
\frac{1}{\sqrt{2\pi \mathrm{Var}[L]}}
\exp\left[-\frac{\left(L_t^{\mathrm{data}} - \mathrm{E}[L]\right)^2}{2\mathrm{Var}[L]}\right],
\end{equation}
where $1\leq t \leq T$ is the trial number. Due to the complicated
functional form of the mean and variance of $L$, creating a maximum
likelihood scheme to estimate $\mu$ from experimental data is
intractable, so we use Eq.~\ref{EQ_LUC_EXP} by replacing the expected
value with the experimental average of measurements
$L_t^{\mathrm{data}}$. If we assume no cooperative transcription
binding ($h=1$), we solve for the estimate $\hat{\mu}$ by applying the
Newton-Raphson iterative method to the equation
\begin{equation}\label{EQ_LUC_NEWTON}
0= \frac{1}{T}\sum_{t=0}^T L_t^{\mathrm{data}} - 
L_0f_{\mathrm{max}}Me^{-\hat{\mu}}\sum_{r=0}^{N_0}\frac{r \hat{\mu}^r}{(K + r)r!}.
\end{equation}
The typical method, under the assumption that luminescent intensity is
proportional to the number of IUs $N$, is to use the sample mean via
the formula $\hat{\mu}^{\mathrm{init}}= \frac{1}{L_0MT}\sum_{t=0}^T
L_t^{\mathrm{data}}$. 
\begin{figure}[t!]
\begin{center}
\includegraphics[width=2.5in]{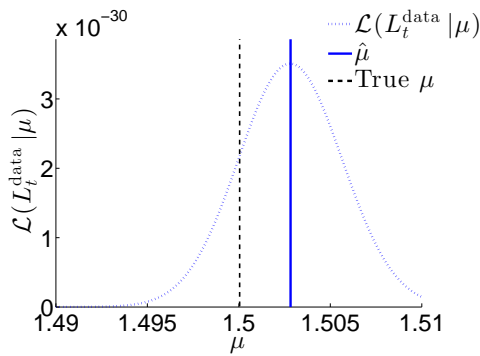}
\caption{The likelihood function $\mathcal{L}(L_t^{\mathrm{data}}
  \vert \mu)$ using Eq.~\ref{EQ_LUC_LIKELIHOOD} and simulated data. We
  set $\mu=1.5$ and assign other parameters with $M=10^5$,
  $f_{\mathrm{max}} = 2$, $h=1$, $K=1$ and $L_0 = 1$. The estimate
  derived from solving Eq.~\ref{EQ_LUC_NEWTON} is $\hat{\mu} = 1.502$
  while the standard method based on the sample mean yields $\hat{\mu}
  = 0.97$, far lower than what is displayed in the plot.}
\label{FIG_SIM_LUC}
\end{center}
\end{figure}
This standard approach fails to account for the
effects of SMOI, but can be used to generate an initial guess for
solving Eq.~\ref{EQ_LUC_NEWTON} iteratively. In order to compare the
two estimates, we simulate data similar the descriptions in the
previous two sections. Here, we do not dilute the initial particle
count and, after distributing the $N$ IUs to the $M$ cells with equal
probability, we compile the SMOI configuration and calculate
$L_t^{\mathrm{data}}$ using Eq.~\ref{EQ_LUC}. The results are shown in
Fig.~\ref{FIG_SIM_LUC}. The iterative method produces an estimate
$\hat{\mu}$ far closer to the true value of $\mu$ than the former
method. A similar approach can be used to compare methods for
alternative functional forms of the viral protein transcription
dynamics described in Eq.~\ref{EQ_HILL}.
\begin{table*}[t!]
\caption{A summary of the analytically derived expressions used to
  analyze experimental results. For virus quantification assays, such
  as the plaque and endpoint dilution assays, one typically wishes to
  estimate the number of initial viral particles $N_0$. For luciferase
  reporter infectivity assay, the ratio $\mu=N/M$ is desired. Our
  improved parameter estimation methods are listed next to standard
  methods currently used.}
\label{TABLE_DISCUSSION}
\begin{tabular}{ |p{1.5cm}||p{7.4cm}|p{7.4cm}|  }
 \hline
 \multicolumn{3}{|c|}{Comparison of Virological Assay Analyses} \\
 \hline
 Assay\newline (Parameter) & Standard Method & New Method \\
 \hline
 \hline
 Plaque \newline($N_0$)  
 & ~\newline $\displaystyle \hat{N}_0 = D^{d_c}\left(\frac{1}{T} \sum_{t = 1}^T P_{d_c,t}\right)$  
 & $\displaystyle 0 = \sum_{d = d_c}^{d_{\mathrm{max}}} \sum_{t = 1}^T \frac{M \exp\left(-\frac{\hat{N}_0}{QMD^d}\right) - M + P_{d,t}}{QMD^d\left[1 - \exp\left(-\frac{\hat{N}_0}{QMD^d}\right)\right]}$\newline Initial guess:\newline $\displaystyle \hat{N}_0^{\mathrm{init}} = -QMD^d \ln\left(1 - \frac{1}{MT}\sum_{t = 1}^T P_{d_c, t}\right)$\\
 \hline
 Endpoint\newline Dilution\newline ($N_0$) 
 &  Reed and Muench: \newline $\displaystyle \log_{10}(\hat{N}_0) = d_{50\%} + \frac{E_{d_{50\%}} - 0.5T}{E_{d_{50\%}} - E_{d_{50\%}+1}}$\newline Spearman-Karber: \newline $\displaystyle \log_{10}(\hat{N}_0) = d_{100\%} - \left[\frac{1}{2} - \sum_{d = d_{100\%}}^{d_{\mathrm{max}}}\frac{E_d}{T}\right]\log_{10}D$ 
 & $\displaystyle 0=\sum_{d=1}^{d_{\mathrm{max}}}\frac{E_d - T + T\exp\left(-\frac{\hat{N}_0}{QD^d}\right)}{QD^d\left(1 - \exp\left(-\frac{\hat{N}_0}{QD^d}\right)\right)}$\newline Initial guess:\newline $\displaystyle \hat{N}_0^{\mathrm{init}} = \frac{-Q D^{d_c}}{2}\ln\left[\left(1 - \frac{E_{d_c}}{T}\right)\left(1 - \frac{E_{d_c+1}}{T}\right)^D\right]$\\
 \hline
 Luciferase\newline Reporter  $\left(\mu = \frac{N}{M}\right)$ 
 & ~\newline $\displaystyle \hat{\mu}= \frac{1}{L_0MT}\sum_{t=0}^T L_t^{\mathrm{data}}$ 
 & $\displaystyle 0=\frac{1}{T}\sum_{t=0}^T L_t^{\mathrm{data}} - L_0f_{\mathrm{max}}Me^{-\hat{\mu}}\sum_{r=0}^{N_0}\frac{r \hat{\mu}^r}{(K + r)r!}$\newline Initial guess:\newline $\displaystyle \hat{\mu}^{\mathrm{init}}= \frac{1}{L_0MT}\sum_{t=0}^T L_t^{\mathrm{data}}$\\
 \hline
\end{tabular}
%\end{table}
%\end{tabular}
\end{table*}
%
%\end{widetext}

%%%%%%%%%%%%%%%%%%%%%%%%%%%%%%%%%%%%%%%%%%%%%%%%%%%%%%%
\section{Conclusion}\label{SEC_DISC}
In this work, we derived probability models that quantify the
viral infectivity of host cells in an \textit{in vitro}
environment. By factoring in the stochastic nature of virus-host
engagement, defective and/or abortive events, and the possibility of
multiple infections of a single host, we defined the statistical
multiplicity of infection (SMOI) and determined related probabilistic
models. We analyzed two limiting regimes: small numbers
of infecting viruses $N$ and large $N$. For the low $N$
regime, Eqs.~\ref{EQ_PROB} and \ref{EQ_JOINT_PROB} model how the
limited number of infectious units are distributed amongst the $M$
host cells. Alternatively, for large $N$, we showed the cell counts of
the SMOI become statistically independent, as displayed in
Eq.~\ref{EQ_JOINT_PROB_APPROX}, and that they display a Poisson
distribution (Eq.~\ref{EQ_PROB_APPROX}). Lastly, we explored the
effects of serial dilution on the total number of infected cells and
the probability of observing an infectious signal in
Eq.~\ref{EQ_H_PROB_DIST}.
%\begin{widetext}
%

Using our probability models along with reasonable assumptions of
applied combinatorics and nonlinear inference, we analytically derived
expressions for several virus assays to improve on existing methods of
experimental data analysis. For virus quantification assays, serial
dilution results in low numbers of viral particles. Using the
appropriate probability model, we created new methods of estimating
the particle count $N_0$ in the initial viral stock for the plaque
assay and the endpoint dilution assay. For measuring infectivity of a
viral strain, the objective is to determine the effective multiplicity
of infection $\mu=N/M$ as the ratio of successfully infecting viruses
$N$ and the total number of cells $M$ included in the assay. As these
assays operate under no dilution, we employed the large $N$ limit
probability model to analytically derive expressions for the
luciferase reporter assay to estimate $\mu$. A summary of each
estimation method along with the most commonly used counterpart is
displayed in Table~\ref{TABLE_DISCUSSION}.

VQAs are primarily concerned with inferring $N_0$ and assume \textit{a
  priori} knowledge of $M$ and the particle to PFU ratio $Q$. In
actuality, there can be variability in the number of cells present in
the microtiter well and, as discussed in
Section~\ref{SEC_PROBABILITY}, the true value of $Q$ is dependent on
the particular protocol and particular conditions under which an assay
was performed. If an alternative assay (RNA tagging, spectroscopy,
super-resolution imaging, etc.) not using cell infection can
accurately measure $N_0$, then, in theory, a subsequent infection
assay can be used to infer a more reliable measure of $Q$. In fact, in
our analysis of the plaque assay presented in
Appendix E of the SI, we show that one can
determine a significantly higher amount of information about $Q$ with
the same assay protocol if $N_0$ is \textit{a priori} known, rather
than the reverse case. Thus, one may argue that assays that employ
serial dilution, such as plaque and endpoint dilution assays, may be
better utilized to infer $Q$. Because the underlying likelihood of the
data in all assays would be the same, the same derivation techniques
would follow with respect to $Q$ in order to formulate its maximum
likelihood estimate. This analysis shows the robust utility of a full
probabilistic model and data likelihood function.

Although the derived assay models provide explicit equations for
inference, many of the expressions are analytically unsolvable and
require numerical solutions. To improve the accessibility of some of
our results, we have created a web-based tool (available at
https://bamistry.github.io/SMOI/) that can accept data from either
plaque, endpoint dilution, or luciferase reporter assays and
automatically estimate the parameter of interest. Ultimately, these
tools can be used for analysis of future virological studies, but may
also be useful when revisiting results of studies that stress
quantifying viral infectivity \cite{Johnston,Mistry2016}.  For studies
that use serial dilution assays, our approach stresses the advantages
of using information in the data associated with \textit{all} dilution
numbers rather than just that of the critical dilution.

Our probabilistic models of viral infection can be further generalized
to include, for example, the effects of cell size inhomogeneity,
coinfection, and viral interference. In the Supplemental Information,
we provide a framework that would allow one to explore how these
confounding factors can further alter the signal of a virus
assay. Future refinement of these extensions can help to ultimately
derive a mechanistic model for the probability of a single virus
successfully infecting a host cell, which we defined as
$Q^{-1}$. Understanding this probability of infection can help aid
further experimental design and allow better quantification and
resolution of the infection dynamics of particular viral strains.

%%%%%%%%%%%%%%%%%%%%%%%%%%%%%%%%%%%%%%%%%%%%%%%%%%%%%%%
\section*{Author Contributions}
BM derived mathematical formulae, developed statistical inference
framework, performed simulations, generated plots, and wrote the initial
draft. MRD and TC verified the mathematical results,
contributed to their analyses, and edited the manuscript. TC 
conceptualized, designed, and supervised the research.

\section*{Acknowledgments}
This work was supported in part by grants from the NSF (DMS-1516675)
and the Army Research Office (W911NF-14-1-0472). We are especially
grateful to Dr. Nicholas Webb, Prof. Benhur Lee, and Prof. Jerome Zack
for insightful discussions.

%B.M. derived the mathematical models for SMOI and the subsequent
%statistical inference methods for the three assays
%discussed. B.M. programmed and ran computer simulations in order to
%generate assay data to compare methods. As first author, B.M. provided
%the majority of the writing for the manuscript. M.D. derived general
%results for the SMOI expected value and variance. M.D. heavily edited
%the manuscript and provided consultation on the mathematics and
%theory. T.C. acted as the principle investigator during the
%development of the mathematical models and their
%verification. T.C. also heavily edited the manuscript and provided
%mathematical insight.
%%%%%%%%%%%%%%%%%%%%%%%%%%%%%%%%%%%%%%%%%%%%%%%%%%%%%%%
%\bibliography{MultiplicityOfInfection}
%\bibliographystyle{biophysj}

%\end{document}

\newpage

%%%%%%%%%%%%%%%%%%%%%%%%%%%%%%%%%%%%%%%%%%%%%%%%%%%%%%%
%\setcounter{page}{1}
\setcounter{equation}{0}
\renewcommand\theequation{S\arabic{equation}}
\appendix

\begin{widetext}
\section*{Supplementary Information}
\section{Mathematical Appendices}\label{SEC_MATH}
\subsection*{SMOI Probability}
To derive Eq.~\ref{EQ_PROB}, we index all cells with $i\in
\{1,\cdots,M\}$ and define $A_i^r$ as the event that cell $i$ is
infected by exactly $r$ IUs. Then, given $N$ IUs across all $M$ cells,
the probability of $A_i^r$ is given by
\begin{equation}\label{EQ_EVENT_PROB}
\mathrm{Pr}(A_i^r\vert M, N) = \binom{N}{r}\left(\frac{1}{M}\right)^{r}\left(1-\frac{1}{M}\right)^{N-r}.
\end{equation}
Since cell sizes are assumed to be homogeneous, the probability in
Eq.~\ref{EQ_EVENT_PROB} is the same for all cells, but the events
$\{A_1^r,\cdots,A_M^r\}$ are not independent as the number of IUs $N$
shared among the $M$ cells is finite. Thus, we use the
inclusion-exclusion principle \cite{Lange2003} to derive
\begin{eqnarray}
\mathrm{Pr}(M_r = m_r\vert M, N) &=& \sum_{j=m_r}^M (-1)^{j-m_r}\binom{j}{m_r}\sum_{\substack{I\subset\{1,\cdots,M\}\\  |I| =j}} \mathrm{Pr}\left(\bigcap_{i \in I} A_i^r\right)\nonumber\\
&=& \sum_{j=m_r}^M (-1)^{j-m_r}\binom{j}{m_r}\binom{M}{j} \mathrm{Pr}\left(\bigcap_{i =1}^j A_i^r\right)\nonumber\\
&=& \sum_{j=m_r}^M (-1)^{j-m_r}\binom{j}{m_r}\binom{M}{j} \binom{N}{r,\cdots,r,(N-rj)}\left[\prod_{i=1}^j\left(\frac{1}{M}\right)^r \right] \left(\frac{M-j}{M}\right)^{N-rj}\nonumber\\
&=& \sum_{j=m_r}^M \binom{j}{m_r}\binom{M}{j} \binom{N}{r,\cdots,r,(N-rj)}\frac{(-1)^{j-m_r} \left(M-j\right)^{N-rj}}{M^N}.
\end{eqnarray}
Note that the inner summation in the first identity above is over
every possible collection of cells of size $j$, but as each cell is
identical, the sum can be reduced to a single joint probability with
the binomial degeneracy $\binom{M}{j}$.

\subsection*{Expected Value and Variance}
For the generalized $c$-th moment $\mathrm{E}\left[M_r^c\right]$ of
the number of cells $M_r$ infected by exactly $r$ viruses, we start
with Eq.~\ref{EQ_PROB} to obtain
\begin{eqnarray}\label{EQ_EXP_CALC}
\mathrm{E}\left[M_r^c\right]  &=&  \sum_{m_r=0}^{M} \sum_{j=m_r}^{M} m_r^c (-1)^{j-m_r}\binom{j}{m_r}\binom{M}{j} \left(\frac{N!}{\left(r!\right)^j \left(N - rj\right)!}\right)\frac{\left(M-j\right)^{N-rj}}{M^N}\nonumber\\
&=&  \sum_{j=0}^{M} \left[\sum_{m_r=0}^{j} m_r^c (-1)^{j-m_r} \binom{j}{m_r} \right] \binom{M}{j}  \left(\frac{N!}{\left(r!\right)^j \left(N - rj\right)!}\right)\frac{\left(M-j\right)^{N-rj}}{M^N}
%&=&  \sum_{j=0}^{M} u(j,c) \binom{M}{j}  \left(\frac{N!}{\left(r!\right)^j \left(N - rj\right)!}\right)\frac{\left(M-j\right)^{N-rj}}{M^N}.\nonumber
\end{eqnarray}
To aid our derivation, we define the function $u(j,c)$ as
\begin{eqnarray}\label{EQ_U_FUNC}
u(j,c) &=& \sum_{m = 0}^{j} m^c (-1)^{j - m}\binom{j}{m}\nonumber\\
%&=& \sum_{m = 1}^{j} m^{c-1} (-1)^{j - m - 1 + 1}\binom{j-1}{m-1}j\nonumber\\
&=& j \sum_{k = 0}^{j-1} (k+1)^{c-1} (-1)^{j - 1 - k}\binom{j-1}{k}\nonumber\\
%&=& j \sum_{k = 0}^{j-1} \sum_{i=0}^{c-1} \binom{c-1}{i} k^i (-1)^{j - 1 - k}\binom{j-1}{k}\nonumber\\
&=& j \sum_{i=0}^{c-1} \binom{c-1}{i} \sum_{k = 0}^{j-1}  k^i (-1)^{j - 1 - k}\binom{j-1}{k}\nonumber\\
&=& j \sum_{i=0}^{c-1} \binom{c-1}{i} u(j-1,i).
\end{eqnarray}
This is a recursive relationship from which we can evaluate any
$u(j,c)$ using all $u(j-1,i)$ such that $0\leq i<c$. We evaluate the
first three cases $u(j,0) = \delta_{0,j}$, $u(j,1)= \delta_{1,j}$, and
$u(j,2) = \delta_{1,j} + 2\delta_{2,j}$, where $\delta_{0,j}$ is the
Kronecker delta operator that returns the value $1$ when the two
subscript arguments are equal and $0$ otherwise. We use the result for
$c=1$ and Eq.~\ref{EQ_EXP_CALC} to calculate the expected value of
$M_r$ as
\begin{eqnarray}
\mathrm{E}\left[M_r\right]  %&=&  \sum_{j=0}^{M} u(j,1) \binom{M}{j}  \left(\frac{N!}{\left(r!\right)^j \left(N - rj\right)!}\right)\frac{\left(M-j\right)^{N-rj}}{M^N}\nonumber\\
&=&  \sum_{j=0}^{M} \delta_{1,j} \binom{M}{j}  \left(\frac{N!}{\left(r!\right)^j \left(N - rj\right)!}\right)\frac{\left(M-j\right)^{N-rj}}{M^N}\nonumber\\
%&=&  \binom{M}{1}  \left(\frac{N!}{\left(r!\right)^1 \left(N - r(1)\right)!}\right)\frac{\left(M-1\right)^{N-r(1)}}{M^N}\nonumber\\
&=&   M\binom{N}{r}  \left(\frac{1}{M}\right)^r\left(1 - \frac{1}{M}\right)^{N-r}.
\end{eqnarray}
We obtain the second moment $\mathrm{E}\left[M_r^2\right]$ using the
same method in order to obtain the variance of $M_r$ as
\begin{eqnarray}
\mathrm{Var}\left[M_r\right] &=& \mathrm{E}\left[M_r^2\right] - \mathrm{E}\left[M_r\right]^2\nonumber\\
&=& M\binom{N}{r}  \left(\frac{1}{M}\right)^r\left(1 - \frac{1}{M}\right)^{N-r} + \frac{M(M-1)N! (M-2)^{N-2r}}{(r!)^2 (N-2r)! M^N} - \frac{M^2 (N!)^2 (M-1)^{2N-2r}}{(r!)^2\left[(N-r)!\right]^2 M^{2N}}.
\end{eqnarray}

\subsection*{Asymptotic Approximation}
For the derivation of Eq.~\ref{EQ_PROB_APPROX}, we take the
mathematical limit $N, M \to \infty$ while keeping the ratio
$\mu=\frac{N}{M}$ fixed and approximate Eq.~\ref{EQ_PROB} as follows:
\begin{eqnarray}
\mathrm{Pr}(M_r = m_r \vert M, N) &=& \sum_{j=m_r}^M \frac{j!M!N!(-1)^{j-m_r} (M-j)^{N-rj}}{m_r!(j-m_r)!j!(M-j)! (N-rj)!(r!)^j M^{N-rj}M^{rj}} \nonumber\\
&=& \frac{1}{m_r!} \sum_{j=m_r}^M \frac{(-1)^{j-m_r}}{(j-m_r)! (r!)^j} \left[M \cdots (M-j+1)\right] \frac{\left[N \cdots (N - rj + 1)\right]}{M^{rj}}  \left(1 - \frac{j}{M}\right)^{N-rj} \nonumber\\
%&=& \frac{1}{m_r!} \sum_{j=m_r}^M \frac{(-1)^{j-m_r}}{(j-m_r)! (r!)^j} \left[M \cdots (M-j+1)\right] \left[\mu \cdots \left(\mu - \frac{rj - 1}{M}\right)\right]  \left(1 - \frac{\mu j}{N}\right)^{N-rj} \nonumber\\
&\approx& \frac{1}{m_r!} \sum_{j=m_r}^M \frac{(-1)^{j-m_r}}{(j-m_r)! (r!)^j} M^j \mu^{jr} e^{-\mu j} \nonumber\\
%&=& \frac{1}{m_r!} \left[\frac{M \mu^r e^{-\mu}}{r!}\right]^{m_r} \sum_{j=0}^{M-m_r} \frac{(-1)^j}{j!} \left[\frac{M \mu^r e^{-\mu}}{r!}\right]^j \nonumber\\
&\approx& \frac{1}{m_r!} \left[\frac{M \mu^r e^{-\mu}}{r!}\right]^{m_r} \exp\left[-\frac{M \mu^r e^{-\mu}}{r!}\right]. 
\end{eqnarray}
Note that, although the first approximation requires $j$ in the
summation to be sufficiently smaller than $M$, any contribution from
the summation for $j$ close to $M$ vanishes due to both the $(j-m_r)!$
term in the denominator and the $\left(1 - \frac{j}{M}\right)^{N-rj}$
term approaching $0$. Under the same large $M, N$ limit, we can derive
an asymptotic approximation of the joint probability distribution by
taking the natural log of both sides of Eq.~\ref{EQ_JOINT_PROB}:
\begin{eqnarray}
\ln  \mathrm{Pr}(M_0 = m_0,\cdots,M_N=m_N) &=& \ln \left( \frac{1}{M^N}\right) + \ln M!  + \ln N!  + \sum_{r=0}^N \ln \left(\frac{1}{m_r! (r!)^{m_r}}\right) \nonumber\\
&\approx& -N \ln M + M\ln(M) - M + N\ln(N) - N + \sum_{r=0}^N \ln \left(\frac{1}{m_r! (r!)^{m_r}}\right) \nonumber\\
%&=& N\ln \left( \frac{N}{M}\right) + M\ln M - Me^{-\mu}e^{\mu} - \mu M + \sum_{r=0}^N \ln \left(\frac{1}{m_r! (r!)^{m_r}}\right) \nonumber\\
&=& \ln \mu \left( \sum_{r=0}^N r m_r \right) + \left( \ln M - \mu\right)\left( \sum_{r=0}^N m_r \right) - Me^{-\mu}\left( \sum_{r=0}^{\infty} \frac{\mu^r}{r!} \right) \nonumber \\
&\: & \hspace{1cm} + \sum_{r=0}^N \ln \left(\frac{1}{m_r! (r!)^{m_r}}\right) \nonumber \\
%&=& \sum_{r=0}^N\left[r m_r \ln \mu + m_r \ln M - m_r \mu - \frac{Me^{-\mu}\mu^r}{r!} + \ln\left(\frac{1}{m_r!(r!)^{m_r}}\right)\right] - \sum_{r=N+1}^{\infty} \frac{Me^{-\mu}\mu^r}{r!} \nonumber\\
&=& \sum_{r=0}^N \ln\left[\frac{\mu^{r m_r}M^{m_r}e^{-m_r\mu}}{m_r!(r!)^{m_r}}\exp\left(-\frac{Me^{-\mu}\mu^r}{r!}\right)\right] - \mathcal{O}\left(\frac{M \mu^N}{N!}\right)\nonumber\\
&\approx& \ln\left[ \prod_{r=0}^N \frac{1}{m_r!} \left[\frac{M\mu^r e^{-\mu}}{r!}\right]^{m_r}\exp\left(-\frac{M\mu^r e^{-\mu}}{r!}\right)\right].
\end{eqnarray}
Since the argument in the right-hand-side of the last approximation is
the same as Eq.~\ref{EQ_PROB_APPROX}, we arrive at the result in
Eq.~\ref{EQ_JOINT_PROB_APPROX}.

\subsection*{Number of Infected Cells}
To derive Eq.~\ref{EQ_H_PROB_DIST}, we first define $N_d$ as the
number of virus particles present in the viral solution after dilution
of a factor of $D^d$. Obtaining $N_d$ is effectively analogous to
taking a volume of the initial viral stock scaled by $D^{-d}$ and
counting the number of particles captured in the volume. Thus, we
expect $N_d$ to be Poisson-distributed with mean $N_0 D^{-d}$ and
discrete probability density function given by
\begin{equation} 
\label{EQ_POISSON_DENSITY}
\mathrm{Pr}\left(N_d = n_d\right\vert N_0) =
\frac{1}{n_d!}\left(\frac{N_0}{D^d}\right)^{n_d}\exp\left(-\displaystyle{\frac{N_0}{D^d}}\right).
\end{equation}
Once $N_d$ is chosen from the above distribution, for a given
``particle to PFU ratio'' $Q$, the number of IUs $N$ follows a
binomial distribution with a probability function similar to
Eq.~\ref{EQ_N_BINOM_PROB}, but with $N_0$ replaced with $N_d$. Note
that, given an SMOI $\{M_0,\cdots,M_N\}$, it is immediate that
$M^*=M-M_0$. Using this modified density of $N$ and Eqs.~\ref{EQ_PROB}
and \ref{EQ_POISSON_DENSITY}, we can derive the discrete probability
density function of $M^*$ at a given dilution number $d$ as
\begin{eqnarray}
\mathrm{Pr}\left(M^* = m\right) &=& \sum_{n_d = 0}^{N_0} \sum_{n = 0}^{n_d} \mathrm{Pr}(N = n\vert N_d = n_d) \mathrm{Pr}(M_0 = M - m\vert N = n) \mathrm{Pr}(N_d = n_d) \nonumber\\
%&=& \sum_{n_d = 0}^{N_0} \sum_{n = 0}^{n_d} \left[\binom{n_d}{n}\left(Q^{-1}\right)^n\left(1 - Q^{-1}\right)^{n_d - n}\right] \left[\sum_{j = M-m}^{M} (-1)^{j - M + m}\binom{j}{M-m}\binom{M}{j}\frac{(M-j)^n}{M^n}\right]  \left[\frac{\left(\frac{N_0}{D^d}\right)^{n_d}e^{-\frac{N_0}{D^d}}}{n_d!}\right]\nonumber\\
%&=& \sum_{j = M-m}^{M} (-1)^{j - M + m} \binom{j}{M-m}\binom{M}{j} e^{-\frac{N_0}{D^d}} \sum_{n_d = 0}^{N_0} \frac{\left(\frac{N_0}{D^d}\right)^{n_d}}{n_d!} \left[\sum_{n = 0}^{n_d} \binom{n_d}{n} \left(1 - \frac{j}{M}\right)^{n}\left(Q^{-1}\right)^n\left(1 - Q^{-1}\right)^{n_d - n}\right]\nonumber\\
&=& \sum_{j = M-m}^{M} (-1)^{j - M + m} \binom{j}{M-m}\binom{M}{j} e^{-\frac{N_0}{D^d}} \sum_{n_d = 0}^{N_0} \frac{\left(\frac{N_0}{D^d}\right)^{n_d}}{n_d!} \left[1 - Q^{-1} + Q^{-1}\left(1 - \frac{j}{M}\right)\right]^{n_d}\nonumber\\
&\approx& \sum_{j = M-m}^{M} (-1)^{j - M + m} \binom{j}{M-m}\binom{M}{j}  \exp\left[\frac{N_0}{D^d}\left(1 - \frac{j}{QM}\right) - \frac{N_0}{D^d}\right]\nonumber\\
%&=&  \binom{M}{m} \exp\left[\frac{-N_0}{QMD^d}(M-m)\right] \sum_{j=0}^{m} (-1)^j \binom{m}{j}\exp\left(-\frac{N_0}{QMD^d}\right)^j(1)^{m-j}\nonumber\\
&=&  \binom{M}{m} \left[1 - \exp\left(-\frac{N_0}{QMD^d}\right)\right]^m \exp\left(-\frac{N_0}{QMD^d}\right)^{M-m}.
\end{eqnarray}
Note that the approximation that closes the exponential term in the final result employs the assumption that $N_0$ is sufficiently large.

%------------------------------------------------------------------------------------------------------------------------------------------------
\section{Inhomogeneous Cell Size}
\label{SEC_CELL_SIZE}
We derived the probability distribution in Eq.~\ref{EQ_PROB} assuming
the plated host cells are of identical size and volume. This may not
necessarily be the case as each cell exists at different stages of the
mitotic cycle, will attach to the plate bottom at random locations,
and contain deformities in shape and size. Assuming cells cover the
entire surface of the well bottom, Pineda et al. \cite{Pineda2004}
showed that the cell size proportion $p_i$ for cell $i$ is gamma
distributed with probability density
\begin{equation}\label{EQ_PROB_SIZE}
f(p_i)=\frac{M^{\nu}\nu^{\nu}p_i^{\nu-1}\exp(-\nu M p_i)}{\Gamma(\nu)},
\end{equation}
where $\nu$ is a parameter that can be estimated, for example, by
fitting imaging data of cells. Under a specific realization of cell
size distributions $\{p_1,\cdots,p_M\}$, we define $A_i^r$ as the
event that cell $i$ is infected by exactly $r$ viruses with
probability
\begin{equation}
\mathrm{Pr}(A_i^r) = \binom{N}{r}p_i^{r}(1-p_i)^{N-r}.
\end{equation}
Using the inclusion-exclusion principle as above, we derive the
conditional probability distribution of the number of cells $M_r$ that
were infected by exactly $r$ viruses as
\begin{eqnarray}\label{EQ_PROB_INHOMO}
\mathrm{Pr}(M_r = m_r \vert p_1, \cdots, p_M) &=& \sum_{j=m_r}^M (-1)^{j-m_r}\binom{j}{m_r}\sum_{\vert \{i_w\}\vert = j} \mathrm{Pr}\left(\bigcap_{w=1}^j A_{i_w}^r\right)\nonumber\\
&=& \sum_{j=m_r}^M (-1)^{j-m_r}\binom{j}{m_r}\sum_{\vert \{i_w\}\vert = j} \binom{N}{r,\cdots,r,(N-rj)}p_{i_1}^r\cdots p_{i_j}^r \left(1-\sum_{w=1}^j p_{i_w}\right)^{N-rj}\nonumber\\
&=& \sum_{j=m_r}^M (-1)^{j-m_r}\binom{j}{m_r}\sum_{\vert \{i_w\}\vert = j} \frac{N!}{(r!)^j(N-rj)!} \left(\prod_{w=1}^j p_{i_w} \right)^r \left(1-\sum_{w=1}^j p_{i_w}\right)^{N-rj}\hspace{-7mm}. 
\end{eqnarray}
In order to obtain the full probability, we first take note that each
cell size proportion $p_i$ is dependent on each other as they are
constrained by $\sum_i^M p_i = 1$. We avoid this dependency by
noticing the expression in Eq.~\ref{EQ_PROB_SIZE} approaches zero very
rapidly as $p_i$ moves away from the expected value $1/M$. If we
define a sufficiently large proportion $\hat{p}$ such that the
interval $[0,\hat{p}]$ contains the majority of the area under the
probability density in Eq.~\ref{EQ_PROB_SIZE}, we can make the
approximation
\begin{eqnarray}
\mathrm{Pr}(M_r = m_r) &=& \int_0^1 \cdots \int_0^1 \mathrm{Pr}(M_r = m_r \vert p_1, \cdots, p_M) f(p_1, \cdots, p_M) \dd p_1 \cdots \dd p_M \nonumber\\
%&\approx& \int_0^{\hat{p}} \cdots \int_0^{\hat{p}} \mathrm{Pr}(M_r = m_r \vert p_1, \cdots, p_M) f(p_1) \cdots f(p_M) \dd p_1 \cdots \dd p_M\nonumber\\
&\approx&  \left[\frac{M^{\nu}\nu^{\nu}e^{-\nu}}{\Gamma(\nu)}\right]^M \int_0^{\hat{p}} \cdots \int_0^{\hat{p}} \mathrm{Pr}(M_r = m_r \vert p_1, \cdots, p_M) \left(\prod_{w=1}^M p_w \right)^{\nu - 1} \dd p_1 \cdots \dd p_M. 
\end{eqnarray}

It is clear that introducing cell size inhomogeneity dramatically
increases the complexity of our probabilistic SMOI model. For
relatively small numbers of cells $M$, image processing can be used to
determine an estimation of a particular realization of cell size
distribution $\{p_1,\cdots,p_M\}$ for a given experiment and factored
into Eq.~\ref{EQ_PROB_INHOMO}. Note that once the probability
distribution of cell counts $\{M_0,\cdots,M_N\}$ is determined for a
given realization of cell sizes $\{p_1,\cdots,p_M\}$, all subsequent
analysis and derivations follow the same way as in the homogeneous
cell size assumption.

%------------------------------------------------------------------------------------------------------------------------------------------------
\section{Coinfection}\label{SEC_COINFECTION}
As a vector for infection, the primary function of a single virus
particle is to deliver its genetic contents into the host cell
cytoplasm or nucleus \cite{Wilen2012,Qian2009,Boulant}. The typical
model for viral infection assumes each virus contains all the genetic
material required to replicate within a host cell
\cite{Chikere201381,Johnston}. Certain plant and fungi viruses,
however, require two or more particles to successfully replicate
within a host cell since each particle contains only part of the
complete genome \cite{Aguilera2017}. Similarly, RNA viruses that
target animal cells undergo error prone replication, resulting in
partially complete genome sequences. These damaged viral genes may
encode proteins needed for the host cell to successfully replicate new
viruses. In this case, regardless of a successful viral infection, new
viruses capable of infecting further host cells will not be
produced. Additional viral infections that contain the missing
sequence fragments, though, can ``rescue'' the cell's ability to
replicate the virus, a phenomenon known as coinfection. In the context
of our definition of SMOI, we now make the distinction between $M_r$,
the number of cells that have been infected by viral genomes from
exactly $r$ distinct virus particles, and $M_r^{*}$, the number of
cells that are fully capable of replicating new functioning viruses
upon undergoing $r$ distinct viral infections. It is immediate that
each $M_r^* \leq M_r$ and their sum $M^*\equiv \sum_{r=1}^N M_{r}\leq
M-M_0$, so the results in Eqs.~\ref{EQ_H_PROB_DIST} and
\ref{EQ_PROB_CYT} are not sufficient to quantify the total number of
virus-producing cells.

In order to model coinfection, we need to consider the genome of the
virus species of interest. Specifically, we assume the genome is made
up of $G$ distinct genes. For example, many variants of HIV-1 carry a
gene sequence containing $G=9$ genes \cite{Wilen2012}. In our model,
we assume each gene encodes a protein that is essential for
replication. Though individual nucleotide changes due to random
mutations may result in an amino acid chain that is no longer
functioning, some genes may be robust to these changes due to codon
degeneracy or the gene's shear length \cite{Stern2014}. Thus, we
assume each gene $g=1,\cdots,G$ contained within a viral particle has
a probability $q_g$ of losing function. If a cell is infected by
exactly $r$ viral genomes, we define $B_g^r$ as the event that gene
$g$ is still no longer functional, so that $\mathrm{Pr}(B_g^r) =
q_g^r$. To quantify the probability that $k$ genes are no longer
functional in a host cell that has been infected by exactly $r$ viral
genomes, we use the inclusion-exclusion principle \cite{Lange2003} to
derive
\begin{eqnarray}\label{EQ_PROB_GENE_FAIL}
\mathrm{Pr}\left(\text{``$k$ failed genes given $r$ infections''}\right) &=& \sum_{j = k}^G (-1)^{j-k} \binom{j}{k}\sum_{\substack{I\subset\{1,\cdots,G\}\\  |I| =j}} \mathrm{Pr}\left(\bigcap_{g \in I} B_g^r\right)\nonumber\\
&=& \sum_{j = k}^G (-1)^{j-k} \binom{j}{k}\sum_{\sigma_1=0}^1\cdots \sum_{\sigma_G=0}^1 \mathbbm{1}_{\sum_{g=1}^G \sigma_g = j}\prod_{g=1}^G q_g^{\sigma_g r},
\end{eqnarray}
where $\mathbbm{1}_{\sum_{g=1}^G \sigma_g = j}$ is an indicator
function that returns zero when the number of nonzero $\sigma_g$ is
not exactly $j$. The infected cell is only capable of producing viable
viruses if none of the genes have failed and is equivalent to setting
$k=0$ in Eq.~\ref{EQ_PROB_GENE_FAIL}. Then we define the probability
$H_r$ that a cell infected by exactly $r$ viral genomes will
successfully produce new viruses as
\begin{equation}
H_r = \sum_{j = 0}^G (-1)^{j} \sum_{\sigma_1=0}^1\cdots \sum_{\sigma_G=0}^1 \mathbbm{1}_{\sum_{g=1}^G \sigma_g = j}\prod_{g=1}^G q_g^{\sigma_g r}.
\end{equation}
Note that the probability that a cell not infected by any viral genome
will produce viruses is $H_0 = 0$. Then, given an SMOI
$\{M_0,\cdots,M_N\}$, the number of cells $M_r^*$ capable of virus
replication after being infected by exactly $r$ viral genomes is
binomially distributed with parameters $M_r$ and $H_r$. The
probability of $M^*$ cells producing viruses is given by
\begin{eqnarray}
\mathrm{Pr}\left(M^* = m\vert M_0,\cdots,M_N,M,N\right) &=& \sum_{M_1^*,\cdots,M_N^*} \binom{m}{M_1^*,\cdots,M_N^*} \prod_{r = 1}^N \binom{M_r}{M_r^*}H_r^{M_r^*}(1 - H_r)^{M_r - M_r^*}.
\end{eqnarray}
If we let $m=0$ and sum over the density in Eq.~\ref{EQ_JOINT_PROB}
for all possible SMOI, given an IU count $N$, we can derive the
probability of observing a cytopathic effect as
\begin{eqnarray}
\mathrm{Pr}(\text{``Cytopathic effect''}\vert N) %&=& 1 - \mathrm{Pr}\left(M^* = 0\vert N\right)\nonumber\\ 
&=& 1 - \sum_{M_0,\cdots,M_N} \frac{1}{M^N}\binom{M}{M_0,\cdots,M_N} \binom{N}{0,\cdots,0,1,\cdots,1,\cdots,N,\cdots,N} \prod_{r = 1}^N (1 - H_r)^{M_r}\nonumber\\
&=& 1 - \frac{M! N!}{M^N} \prod_{r=0}^N \sum_{M_r = 0}^M \frac{(1-H_r)^{M_r}}{M_r! (r!)^{M_r}}\nonumber\\
%&\approx& 1 - \frac{M! N!}{M^N} \prod_{r=0}^N \exp\left[ \frac{1-H_r}{r!}\right]\nonumber\\
&\approx& 1 - \frac{M! N!}{M^N}  \exp\left[\sum_{r = 0}^N \frac{1-H_r}{r!}\right],
\end{eqnarray}
where the approximation is due to the assumption that the number of
cells $M$ is large. For intermediate values of $N$, computing the
summation in the exponential is numerically viable, assuming the
probabilities of gene failure $q_1,\cdots, q_G$ are known. Though this
expression may be used in place of Eq.~\ref{EQ_PROB_CYT} to analyze
some virus quantification assays, for large values of $N$, numerically
evaluating $H_r$ becomes computationally expensive.

%--------------------------------------------------------------------------------------------------------------------------------------------------
\section{Viral Interference}\label{SEC_INTERFERENCE}
To infect healthy cells, all species of viruses must undergo a series
of events including cell attachment, entry via membrane fusion or
endocytosis, and intracellular transport. Retroviruses, such as HIV-1,
must also undergo reverse transcription, nuclear pore transport, and
DNA integration in order to use the host cell's transcription
machinery to produce viral protein. In the models developed in this
paper, the probabilities of success for each of these processes was
assumed to be subsumed into the \textit{a priori} estimated particle
to PFU ratio $Q$. However, for certain retroviruses, it has been
observed that after an initial infection, subsequent infections from
the same virus species become less likely \cite{Nisole2004,
  Nethe2005}. This phenomenon, known as viral interference, is often
due to the host producing new viral proteins after a refractory period
that can inhibit one or more of the intracellular processes leading to
integration of subsequent viral infections. To include this dynamic
into our models, we first decouple the probabilities of integration
from $Q$ and define $N$ as the number of viruses that have
successfully completed viral entry into the host cytoplasm, but before
all intracellular processes that lead to integration. Note that all of
our results concerning the statistical multiplicity of infection
(SMOI) still hold and we make the distinction between the number $M_r$
of cells infected by $r$ of the $N$ infectious units and the number
$M_s^*$ of cells with exactly $s$ integrations. Furthermore, some
species of virus can contain multiple copies of their genome, such as
HIV-1 which contains two copies per particle \cite{Wilen2012}. Let $a$
be the number of genomes contained in a single virus particle to be
integrated into the host cell. Then the maximum number of possible
integrations for a cell from $M_r$ is $ra$. Let $p_s$ be the
probability of a viral genome integrating into the host DNA given that
$s-1$ integrations have already occurred. Define $H_{r,s}$ as the
probability a cell contains $s$ successful integrations given that it
was infected by exactly $r$ distinct virus particles and is given by
\begin{equation}
H_{r,s}= \begin{cases} 
      p_1 p_2 \cdots p_s (1 - p_{s+1})^{ra-s} & 0 \leq s\leq ra \\
      0 & s > ra.
   \end{cases}
\end{equation}
If we define $M_{r,s}^*$ as the number of cells with $s$ integrations
after infection by exactly $r$ virus particles, then given an SMOI
$\{M_0,\cdots,M_N\}$ and $N$, we can derive the probability function
\begin{equation}\label{EQ_PROB_INTEGRATION}
\mathrm{Pr}(M_{r,s}^* = m \vert M_0,\cdots,M_N,N) = \binom{M_r}{m}H_{r,s}^m \left(1 - H_{r,s}\right)^{M_r - m}.
\end{equation}
Noting that $M_s^*=\sum_{r = 0}^N M_{r,s}^*$ is the number of cells
with exactly $s$ integrations, we can use Eqs.~\ref{EQ_PROB_APPROX}
and \ref{EQ_PROB_INTEGRATION} to derive the expected value as
\begin{align}\label{EQ_EXP_INTEGRATION}
\mathrm{E}\left[M_s^*\vert N\right] &= \sum_{r = 0}^N \mathrm{E}\left[M_{r,s}^*\vert N\right]\nonumber\\
&= \sum_{r = 0}^N H_{r,s}\mathrm{E}\left[M_r\vert N\right]\nonumber\\
&= M e^{-\mu} \sum_{r = 0}^N \frac{H_{r,s}\mu^r}{r!},
\end{align}
where $\mu = \frac{N}{M}$. Note that if we are concerned with the
total number $M^*=M - M_0^*$ of cells with at least one integration,
as is the case for the probability distributions derived for assays
employing serial dilution, issue of viral interference is negligible,
allowing us to subsume the probability of the first integration into
the particle to PFU ratio $Q$ as before and leave all subsequent virus
quantification analysis unchanged from the results in
Section~\ref{SEC_PLAQUE} and \ref{SEC_ENDPOINT}. However, for assays
that attempts to quantify the total number of integrations, such as
the luciferase reporter assay, the expectation in
Eq.~\ref{EQ_EXP_INTEGRATION} can be used, assuming the probabilities
$p_1,\cdots,p_N$ have \textit{a priori} been estimated.

%--------------------------------------------------------------------------------------------------------------------------------------------------
\section{Sensitivity Analysis}
\label{SEC_SENSITIVITY}
The probability models derived in Section~\ref{SEC_PROBABILITY}
allowed us to construct the likelihood functions for the plaque,
endpoint dilution, and luciferase reporter assays in
Eqs.~\ref{EQ_PLAQUE_LIKELIHOOD}, \ref{EQ_EDA_LIKELIHOOD}, and
\ref{EQ_LUC_LIKELIHOOD} for the primary purpose of inferring unknown
parameters such as $N_0$ and $\mu$. The utility of these functions can
be extended to performing sensitivity analysis on these maximum
likelihood estimates (MLE) and optimizing experimental design. This
requires constructing a Fisher Information Matrix (FIM), a
quantitative measure of the information one can extract for a
likelihood function with an arbitrary set of data \cite{Gunawan2005,
  Casella2002}. The FIM, which we will denote as $J$, is constructed
by computing the gradient of the log of the likelihood function with
respect to the parameters being inferred. For example, for the plaque
assay and potentially inferred parameters $N_0$, $Q$, and $M$, $J$ is
given by
\begin{equation}
J = \mathrm{E}\left[\left(\nabla \ln \mathcal{L}\right)\left(\nabla \ln \mathcal{L}\right)^{\mathrm{T}}\right] = \begin{bmatrix}
J_{N_0,N_0}& J_{N_0,Q}& J_{N_0,M}\\
J_{Q,N_0}& J_{Q,Q}& J_{Q,M}\\
J_{M,N_0}& J_{M,Q}& J_{M,M}\\
\end{bmatrix},
\end{equation}
where we derive
\begin{align}
J_{N_0,N_0} &= \mathrm{E}\left[\left(\frac{\partial \ln \mathcal{L}}{\partial N_0}\right)^2\right] =  \sum_{d = d_{\mathrm{c}}}^{d_{\mathrm{max}}} \frac{T \exp\left(-\frac{N_0}{Q M D^d}\right)}{Q^2 M D^{2d}\left[1 - \exp\left(-\frac{N_0}{Q M D^d}\right)\right]},\label{EQ_FIM_PLAQUE}\\
J_{Q,Q} &= \mathrm{E}\left[\left(\frac{\partial \ln \mathcal{L}}{\partial Q}\right)^2\right] = \sum_{d = d_{\mathrm{c}}}^{d_{\mathrm{max}}} \frac{T N_0^2 \exp\left(-\frac{N_0}{Q M D^d}\right)}{Q^4 M D^{2d}\left[1 - \exp\left(-\frac{N_0}{Q M D^d}\right)\right]},\\
J_{M,M} &= \mathrm{E}\left[\left(\frac{\partial \ln \mathcal{L}}{\partial M}\right)^2\right] = \sum_{d = d_{\mathrm{c}}}^{d_{\mathrm{max}}} \frac{T N_0 \exp\left(-\frac{N_0}{Q M D^d}\right)}{Q M^2 D^{d}\left[1 - \exp\left(-\frac{N_0}{Q M D^d}\right)\right]},\\
J_{N_0,Q} &= J_{Q, N_0} = \mathrm{E}\left[\left(\frac{\partial \ln \mathcal{L}}{\partial N_0}\right) \left(\frac{\partial \ln \mathcal{L}}{\partial Q}\right)\right] = -\sum_{d = d_{\mathrm{c}}}^{d_{\mathrm{max}}} \frac{T N_0 \exp\left(-\frac{N_0}{Q M D^d}\right)}{Q^3 M D^{2d}\left[1 - \exp\left(-\frac{N_0}{Q M D^d}\right)\right]},\\
J_{N_0,M} &= J_{M, N_0} = \mathrm{E}\left[\left(\frac{\partial \ln \mathcal{L}}{\partial N_0}\right) \left(\frac{\partial \ln \mathcal{L}}{\partial M}\right)\right] = -\sum_{d = d_{\mathrm{c}}}^{d_{\mathrm{max}}} \frac{T N_0 \exp\left(-\frac{N_0}{Q M D^d}\right)}{Q^2 M^2 D^{2d}\left[1 - \exp\left(-\frac{N_0}{Q M D^d}\right)\right]},\\
J_{Q,M} &= J_{M,Q} = \mathrm{E}\left[\left(\frac{\partial \ln \mathcal{L}}{\partial Q}\right) \left(\frac{\partial \ln \mathcal{L}}{\partial M}\right)\right] = \sum_{d = d_{\mathrm{c}}}^{d_{\mathrm{max}}} \frac{T N_0^2 \exp\left(-\frac{N_0}{Q M D^d}\right)}{Q^3 M^3 D^{2d}\left[1 - \exp\left(-\frac{N_0}{Q M D^d}\right)\right]}.
\end{align}
In particular, the elements of the main diagonal of $J$, known as
Fisher Information Numbers, are interpreted as the ``precision'' of
each MLE and can inform an experimentalist of the potential variation
in their inferred parameter with respect to data defined by the
likelihood function. Comparing the main diagonal elements can offer
insight into experimental design. To illustrate, in the example above,
it is immediately apparent that the ratio of $J_{Q,Q}$ to
$J_{N_0,N_0}$ is $N_0^2/Q^2$, where it is understood that $N_0$ is
typically several orders of magnitude higher than $Q$. This implies
that the likelihood function of Eq.~\ref{EQ_PLAQUE_LIKELIHOOD}, and,
by extension, the plaque assay itself contains far more information
about the parameter $Q$ than $N_0$. This provides an analytical way to
decide which parameter estimation should be the focus of a particular
assay.

A more general use for the FIM is to understand the variance of an MLE
given an arbitrary set of data. Independent, but identical experiments
can produce different estimates for each parameter and, according to
the Cramer-Rao inequality, the matrix inverse $J^{-1}$ will provide a
theoretical lower bound on the covariance matrix of the parameter
estimates \cite{Gunawan2005}. Furthermore, it can be shown that the
distribution of MLEs asymptotically approaches a normal distribution
centered around the true experimental parameter value with covariance
$J^{-1}$ as the amount of data increases \cite{Sobel1982}. For single
point estimation, the FIM reduces to the one Fisher Information Number
with which the reciprocal can be used to approximate the variance of a
parameter. For example, the plaque assay is typically used to infer
only the parameter $N_0$, so using Eq.~\ref{EQ_FIM_PLAQUE}, we can
obtain the asymptotic approximation
\begin{equation}
\mathrm{Var}\left[\hat{N}_0\right] \approx J_{N_0,N_0}^{-1} = \left[\sum_{d = d_{\mathrm{c}}}^{d_{\mathrm{max}}} \frac{T \exp\left(-\frac{N_0}{Q M D^d}\right)}{Q^2 M D^{2d}\left[1 - \exp\left(-\frac{N_0}{Q M D^d}\right)\right]}\right]^{-1}.
\end{equation} 
This analytical expression for the variance can be used to determine
confidence intervals of the MLE, perform sensitivity analysis of other
parameters, and aid in optimal experimental design.

\end{widetext}

\bibliography{MultiplicityOfInfection}
\bibliographystyle{biophysj}

\end{document}